\documentclass[preprint2,longabstract]{aastex}

\shorttitle{Brown Dwarf Candidates in M\,4}
\shortauthors{Dieball et al.}

\begin{document}

\title{Deep near-IR observations of the Globular Cluster M\,4: Hunting
  for Brown Dwarfs.}

\author{A. Dieball\altaffilmark{1}}
\affil{Argelander Institut f\"ur Astronomie, Helmholtz Institut f\"ur Strahlen- und Kernphysik, University of Bonn, Germany}
\email{adieball@astro.uni-bonn.de}

\author{L. R. Bedin}
\affil{INAF–-Osservatorio Astronomico di Padova, Vicolo dell'Osservatorio 5, I-35122 Padova, Italy}

\author{C. Knigge}
\affil{Physics and Astronomy, University of Southampton, SO17 1BJ, UK}

\author{R. M. Rich}
\affil{Department of Physics and Astronomy, University of California at Los Angeles, Los Angeles, CA 90095-1562, USA}

\author{F. Allard}
\affil{Centre de Recherche Astrophysique de Lyon, UMR 5574: CNRS, Université de Lyon, École Normale Supérieure de Lyon, 46 allée d'Italie, 69364 Lyon Cedex 07, France}

\author{A. Dotter}
\affil{Research School of Astronomy and Astrophysics, Australian National University, Canberra, ACT, Australia}

\author{H. Richer}
\affil{Department of Physics and Astronomy, University of British Columbia, Vancouver, BC, V6T 1Z1, Canada }

\author{D. Zurek}
\affil{Department of Astrophysics, American Museum of Natural History, New York, NY 10024, USA}

\altaffiltext{1}{Visiting Astronomer, Physics and Astronomy, University of Southampton, SO17 1BJ, UK}

\begin{abstract}
  We present an analysis of deep {\it HST}/WFC3 near-IR (NIR) imaging
  data of the globular cluster M\,4. The best-photometry NIR
  colour-magnitude diagram (CMD) clearly shows the main sequence
  extending towards the expected end of the Hydrogen-burning limit and
  going beyond this point towards fainter sources. The white dwarf
  sequence can be identified. As such, this is the deepest NIR CMD of
  a globular cluster to date. Archival {\it HST} optical data were
  used for proper-motion cleaning of the CMD and for distinguishing
  the white dwarfs (WDs) from brown dwarf (BD) candidates. Detection
  limits in the NIR are around $F110W\approx26.5$ mag and
  $F160W\approx27$ mag, and in the optical around $F775W\approx28$
  mag. Comparing our observed CMDs with theoretical models, we
  conclude that we have reached beyond the H-burning limit in our NIR
  CMD and are probably just above or around this limit in our
  optical-NIR CMDs. Thus, any faint NIR sources that have no optical
  counterpart are potential BD candidates, since the optical data are
  not deep enough to detect them. We visually inspected the positions
  of NIR sources which are fainter than the H-burning limit in $F110W$
  and for which the optical photometry did not return a counterpart.
  We found in total five sources for which we did not get an optical
  measurement. For four of these five sources, a faint optical
  counterpart could be visually identified, and an upper optical
  magnitude was estimated. Based on these upper optical magnitude
  limits, we conclude that one source is likely a WD, one source could
  either be a WD or BD candidate, and the remaining two sources agree
  with being BD candidates. For only one source no optical counterpart
  could be detected, which makes this source a good BD candidate. We
  conclude that we found in total four good BD candidates.
\end{abstract}


\section{Introduction}
\label{intro}

Globular clusters (GCs) are the oldest and most massive stellar
aggregates in our Galaxy. As such, they are the best natural
laboratories to study large, co-eval populations of stars at known
distance and metallicity. Indeed, much of our understanding of star
formation and evolution has been derived from observational studies of
GCs. Nonetheless, we still lack an understanding of the very low mass
stars (VLMSs) around the faint end of the Hydrogen-burning main
sequence (MS) and of objects beyond that limit, i.e., sub-stellar
sources, so called ``Brown Dwarfs'' (BDs). This is especially true for
objects at the low-metallicities typical for the GCs in our Milky Way
(see below).

BDs present a link between stars and planets, and thus are important
for our understanding of both star and planet formation and evolution.
BDs are sub-stellar objects that are not massive enough to ignite and
sustain Hydrogen burning. Thus, where low-mass stars will retain their
luminosity for a Hubble time or longer, a BD will continue to cool and
become fainter with age. Like giant gas planets, BDs have complex
atmospheres \citep{burrows1997, burrows2001}. The distinction between
stars, BDs and planets is either based on mass or on formation. In
general, stars have masses $> 80 M_J$ and can sustain Hydrogen
burning, BDs have masses between 80 and 13 $M_J$ and cannot sustain
Hydrogen burning but a short period of Deuterium burning, and giant
planets have masses below 13 $M_J$ and cannot sustain Deuterium
burning (Burrows et al.\ 1997, Stamatellos 2014, but see also
Sect.~\ref{Hmass}). Because low-mass stars and BDs have life-times
much longer than the age of the Galaxy, the GC VLMSs and BDs are also
important tracers of Galactic formation and chemical evolution.

The formation of BDs is a matter of considerable dispute. They might
have formed in the same way as (low-mass) stars from turbulent cloud
fragmentation \citep{elmegreen1999, wg2005, andre2012}, which would
imply a continuous extension of the IMF into the sub-stellar regime.  On
the other hand, BDs might form e.g. from the ejection of stellar
embryos or sub-stellar clumps which did not have the chance to
accumulate enough mass \citep{rc2001, bv2012}. Kroupa \& Bouvier
(2003) suggested that BDs form via photo-evaporation of protostars
through nearby massive stars. This might also suggest an increase in
the number of BDs with cluster mass, as more massive clusters have
more O stars which can produce more BDs. BDs might also form from the
fragmentation of circumstellar disks (Stamatellos et al.\ 2011, Thies
et al.\ 2010, Kaplan et al.\ 2012). In this case, the number of BDs in
clusters could be enhanced in dense clusters as dynamical interactions
between cluster stars lead to more disk fragmentation. Thies et
al. (2015) concluded that BDs likely form not just via one formation
scenario, but from a combination of various channels.

Large surveys undertaken in the past decade have detected large
numbers of BDs. For example, the Two Micron All Sky Survey (2MASS;
Skrutskie et al.\ 2006), the Sloan Digital Sky Survey (SDSS; York et
al.\ 2000), the United Kingdom Infrared telescope Deep Sky Survey
(UKIDSS; Lawrence et al.\ 2007), the Wide-field Infrared Survey
Explorer (WISE, Wright et al.\ 2010) all sky survey have been very
successful in finding such cool, low-mass
objects.\footnote{Compilations of BDs and low-mass stars can be found
  on DwarfArchives.org, the ``List of Brown Dwarfs'' maintained by
  W. R. Johnston on
  http://www.johnstonsarchive.net/astro/browndwarflist.html, and the
  ``List of all ultracool dwarfs'' maintained by J. Gagne on
  https://jgagneastro.wordpress.com/list-of-ultracool-dwarfs/}
However, for most of these BDs, key physical properties like
metallicity and age are unconstrained (see e.g.\
\mbox{DwarfArchives.org}). In fact, determining the physical parameters is
extremely difficult and a major hurdle in BD research. Observations of
open star clusters and star forming regions, where all sources are at
the same distance and metallicity, can mitigate this problem
(e.g. Steele et al.\ 1995, Rebolo et al.\ 1996, Mart\'{i}n et al.\
2001, Pinfield et al.\ 2003, Boudreault \& Lodieu 2013, Casewell et
al.\ 2014), but only for young and metal-rich objects. Thus the need
for benchmark sources is especially evident for the metal-poor regime,
and indeed we still do not know much about {\it old, metal-poor}
BDs. So far, only very few low-metallicity, old VLMSs near the
H-burning limit, and even fewer sub-stellar (candidate) halo objects
have been identified (e.g. Burgasser et al.~2003, 2009, L\'{e}pine et
al.\ 2004, Burgasser \& Kirkpatrick 2006, Cushing et al.\ 2009,
Sivarani et al.\ 2009, Murray et al.\ 2011, Mace et al.\ 2013,
Pinfield et al.\ 2014, Luhman \& Sheppard 2014, Burningham et al.\
2014, Kirkpatrick et al.\ 2014).

This is where GCs come in: they are massive, and thus might have
produced BDs in large numbers, and they are also the oldest and most
metal-poor stellar aggregates in our Galaxy. Potentially, GCs are the
ideal hunting ground for old, metal-poor benchmark VLMSs and BDs which
are much needed if we are to test stellar and sub-stellar formation
and evolution theories and models of metal-poor (sub-)stellar
atmospheres.

However, identifying substellar objects in GC is challenging due to
their intrinsic faintness. Therefore, the closest GCs make the best
targets for this kind of research. Out of the GCs in our Galaxy, M\,4
(NGC\,6121) is the closest GC to us; distance estimates range from 1.7
kpc (Hansen et al.\ 2004) to $\approx$ 2 kpc (e.g. Bedin et al.\
2009). Braga et al.\ (2015) estimated a true distance modulus of 11.28
mag based on RR Lyrae period-luminosity and period-Wesenheit
relations, resulting in a distance of 1.8 kpc, which agrees well with
previous estimates. Malavolta et al.\ (2014) analyzed 7250 spectra for
2771 cluster stars and found a metallicity of $\rm[Fe/H]=-1.07$ dex
(RGB) to $\rm[Fe/H]=-1.16$ dex (sub giant branch and MS stars), which
is well in line with previous metallicity estimates
(e.g. $\rm[Fe/H]=-1.15$ dex according to the 2010 update of the
Harris, 1996, catalogue of globular clusters in the Milky Way).

As such, M\,4 is a prime target for ultra-deep observational
studies. Indeed, deep optical studies with the Hubble Space Telescope
({\it HST}) have been undertaken by Richer et al.\ (1997, 2004) and
Bedin et al.\ (2009), yielding impressive results.  Richer et
al. (1997, 2004) estimated the fraction of similar-mass photometric
binaries to be small (just 2\% in their outer field, falling to just
1\% towards the cluster core). Milone et al.\ (2012) suggested a much
higher total binary fraction, raising from 10\% at the halfmass radius
to 15\% towards the cluster core. The present day mass function of the
lower-mass MS stars is flat (Bedin et al.\ 2001), with a slope of
$\alpha = 0.1$ and a further flattening towards the cluster centre
(Richer et al.\ 2004). The cluster WDs suggest that the
initial mass function (IMF) above 0.8 $M_\odot$ was much steeper than
the present day mass function. Bedin et al. (2009) presented the
deepest optical colour-magnitude diagram (CMD) to date of this cluster
and located the faint end of the WD cooling sequence in M\,4 at $F606W
= 28.5$ mag, suggesting an age of 11.6$\pm$0.6 Gyr. This agrees with
the finding of Hansen et al. (2004), who found a WD based age of 12.1
Gyr. The ongoing {\it HST} M\,4 core project (PI L. Bedin, GO-12911)
searches for binary dark companions to MS stars. The high-accuracy
astrometry and photometry of this data-set, together with archival
material and a part of the deep near-IR (NIR) data presented in this
paper, have already been used to identify two distinct sequences along
the lower-mass MS \citep{milone2014}.

M\,4 is also so far the only GC known to host a planetary system, PSR
B1620-26 (Sigurdsson et al.\ 2003), which challenged the
planet-metallicity relation of the standard planet formation model at
that time \citep{fischer}. Recently, Hasegawa \& Hirashita (2014)
suggested that the critical metallicity for gas giant formation is
[Fe/H]$\approx −1.2$ dex, which agrees with M\,4's metallicity. Beer
et al.\ (2004) suggested a metallicity-independent formation scenario,
in which the planet in M\,4 formed through dynamically induced
instability in a circumbinary disc. If true, then we can expect many
planets to form especially in the dense GCs in which dynamical
interactions between cluster stars are ubiquitous. Since BDs might
form in a similar way, we then might also expect that many more BDs
form in dense GCs compared to open clusters.

This paper is structured as follows. In Section~\ref{obs} we describe
the observations and reduction of our NIR and archival optical
data. In Section~\ref{cmds} we present and discuss the NIR and
optical-NIR CMDs, and our results and conclusions are summarized in
Section~\ref{summary}.

\section{Observations}
\label{obs}

\subsection{NIR Data}
\label{IRobs}

The NIR observations of the globular cluster M\,4 were carried out in
April 2012 with the Wide Field Camera 3 (WFC3) on board the {\it HST},
using the $F110W$ and $F160W$ filters (program GO--12602, PI:
Dieball). All observations were made at a single pointing position on
a field centred at about one core radius North East from the cluster
centre. This region in the cluster had been the focus of the programs
GO-5461 (Richer et al.\ 1997, 2004) and GO--10146 (Bedin et al.\ 2009)
and thus is fully covered by deep optical observations. A standard
4-point WFC3-IRDITHER-BOX-MIN dither pattern and a sampling of NSAMP14
SPARS50 was applied during the observations to get a well-sampled
point spread function (PSF). The WFC3 field of view is $136\arcsec
\times 123\arcsec$ in the IR channel, with a resolution of
$0.13\arcsec \times 0.121\arcsec$ per pixel. All NIR observations were
carried out during two consecutive orbits on April 16th ($F110W$) and
four orbits on April 20th to 21st ($F160W$), comprising a total of 8
individual images in $F110W$ and 16 images in $F160W$, each 653 seconds,
resulting in total exposure times of 5223 seconds ($F110W$) and 10447
seconds ($F160W$).

Using the pipeline produced flat-fielded (FLT) images, we first
created a master image for each of our IR filters, using {\tt
  multidrizzle} running under {\tt PyRAF} (the Python-based interface
to IRAF\footnote{IRAF (Image Reduction and Analysis Facility) is
  distributed by the National Astronomy and Optical Observatory, which
  is operated by AURA, Inc., under cooperative agreement with the
  National Science Foundation.}). {\tt multidrizzle} corrects
geometric distortions that are present in the input images and
combines them into a master image. Shifts between the individual
images are expected, as we have applied a dither pattern. On top of
this, telescope breathing can affect guide star tracking and as a
result can cause small shifts, typically on a sub-pixel scale (see
e.g.\ the {\tt multidrizzle} handbook available on the STScI
webpages). In order to ascertain that all shifts are taken into
account, we created geometric-distortion corrected individual
images. Based on the coordinates of the same 10 stars in each of these
images (selected to be well distributed over the field of view),
accurate shifts between the individual and the master images were then
determined using {\tt tweakshift}.

\begin{figure*}[h]
\centerline{
\includegraphics[width=\textwidth]{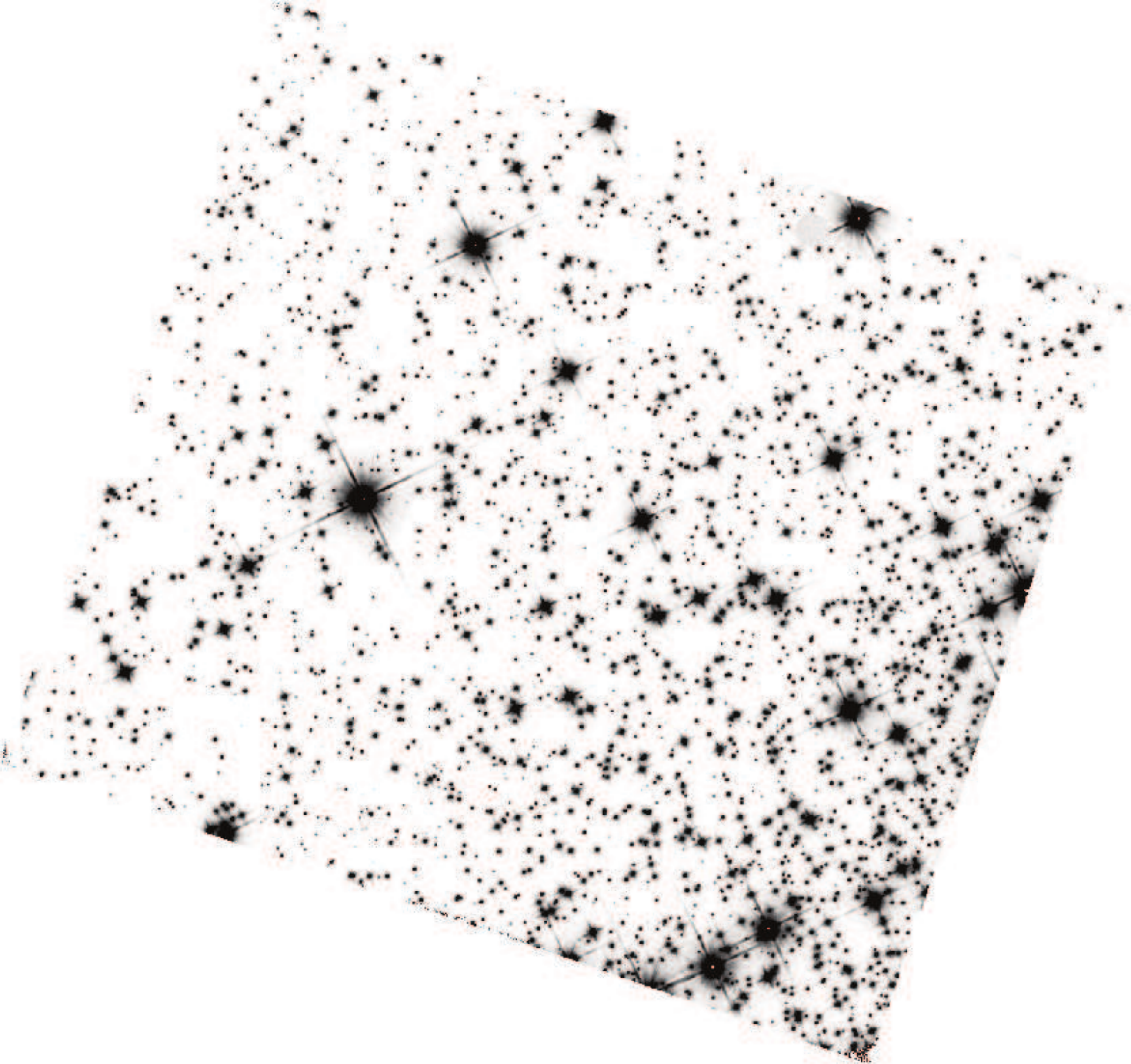}
}
\caption{Left: Geometric-distortion corrected $F110W$ master
  image. North is up and East to the left. The field of view is
  $136\arcsec \times 123\arcsec$. The image is displayed on a
  logarithmic scale to bring out fainter sources.\label{f110}}
\end{figure*}

\begin{figure*}
\centerline{
\includegraphics[width=\textwidth]{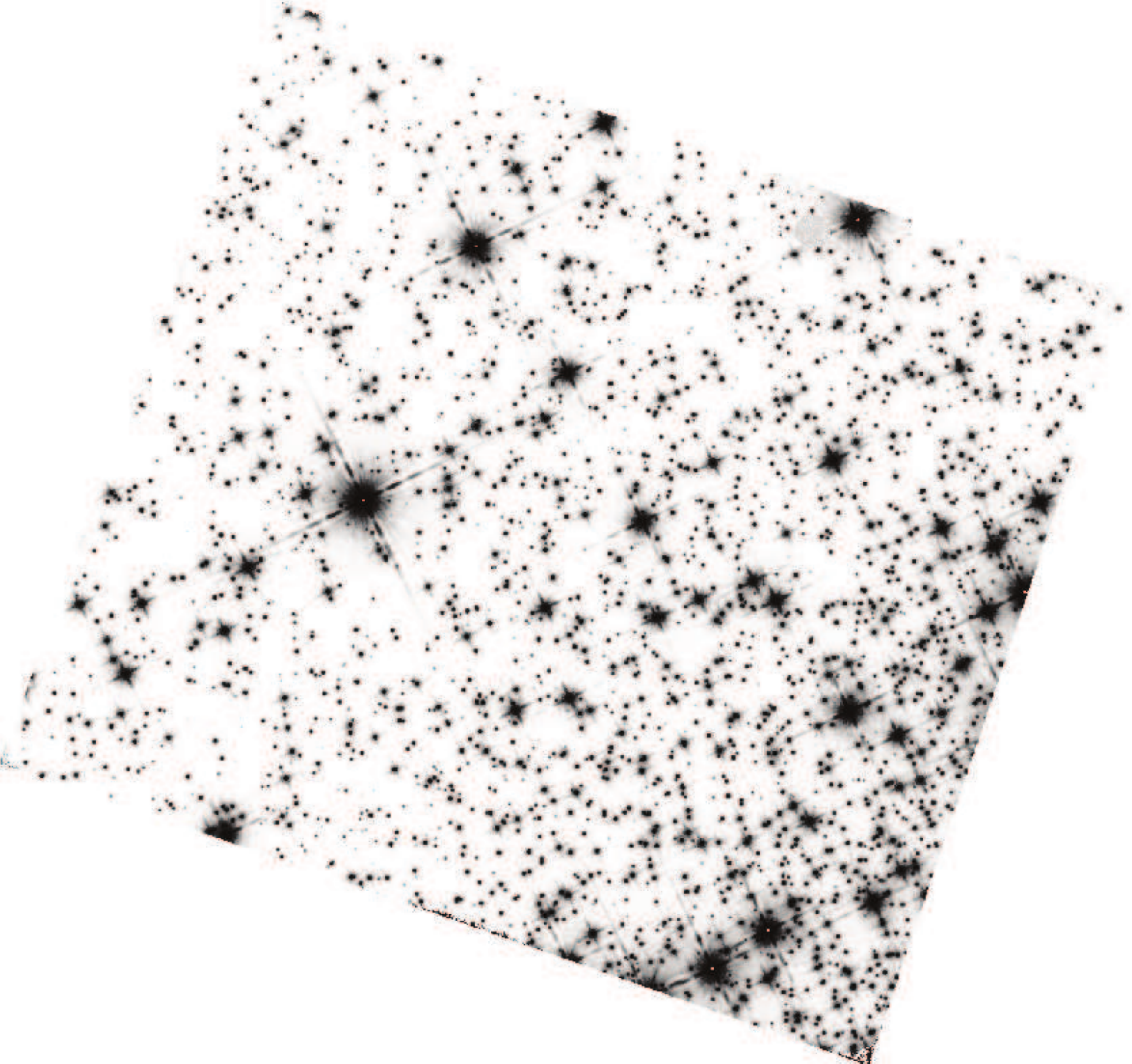}
}
\caption{Same as Fig.~\ref{f110} but for the $F160W$
  data.\label{f160}}
\end{figure*}

The master images created in this way are displayed in
Figures~\ref{f110} and \ref{f160}. Note that these master images serve
as reference images for the positions of the stars detected by {\tt
  DOLPHOT}, but the photometry was actually performed on the
individual images (see \ref{dolphot} below).

\subsection{Photometry of the NIR data}
\label{dolphot}

Photometry was performed on the individual FLT images using the {\tt
  DOLPHOT} software\footnote{http://americano.dolphinsim.com/dolphot/}
developed by A. Dolphin as a generalized version of {\tt HSTphot}
\citep{dolphin2000}. {\tt DOLPHOT} runs on the FLT images downloaded
from the STScI archive, i.e., no further processing of the images is
necessary nor recommended. Photometry is performed on each of the
input images, using the WFC3 module which replaces the analytic PSF
model with a look-up table computed using Tiny Tim PSFs (Krist \& Hook
2011). For our data, we used Jay Anderson's PSF libraries for the
WFC3/IR filters and the pixel area maps available from the {\tt
  DOLPHOT} webpage. The WFC3 module also includes a photometric
calibration to the VEGAmag system. As a reference frame for a common
physical (image) coordinate system we use our deepest image, i.e., the
master image created with {\tt multidrizzle}. The {\tt DOLPHOT}
package also provides routines to mask all pixels that are flagged as
bad in the data quality arrays, to multiply with the pixel area maps,
to calculate sky images, and to align all input images (or rather the
source coordinates found in each input image) to the reference frame
(our drizzled master images) using user defined lists of stellar
coordinates in each input and the reference image. For these steps, we
used the recommended WFC3 IR parameter settings.  

Note that performing photometry on the drizzled image is expected to
provide sub-optimal photometry, because the drizzling process affects
the PSF and the noise characteristics of the drizzled images. Instead,
{\tt DOLPHOT} runs on all input images simultaneously and thus is
capable of providing deep photometry. We started with the recommended
parameters, and then refined some parameters to push our photometry as
deep as possible.\footnote{We used {\tt Force1 = 1}, {\tt FlagMask =
    4} to eliminate saturated stars, {\tt WFC3IRpsfType = 1} for the
  Anderson PSF cores, {\tt FitSky = 2}, {\tt SigFind = 1.5}, {\tt
    SigFindMult = 0.8}, {\tt SigFinal = 1.5}, and {\tt RPSF = 15}.}

\begin{figure*}[h]
\centerline{
\includegraphics[width=7.5cm]{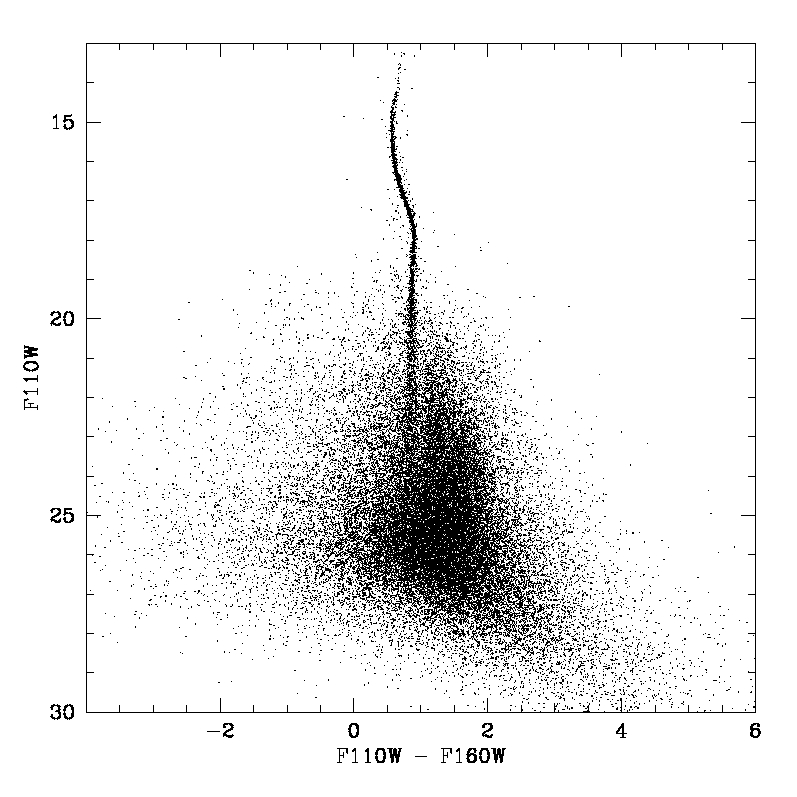}\\
\includegraphics[width=7.5cm]{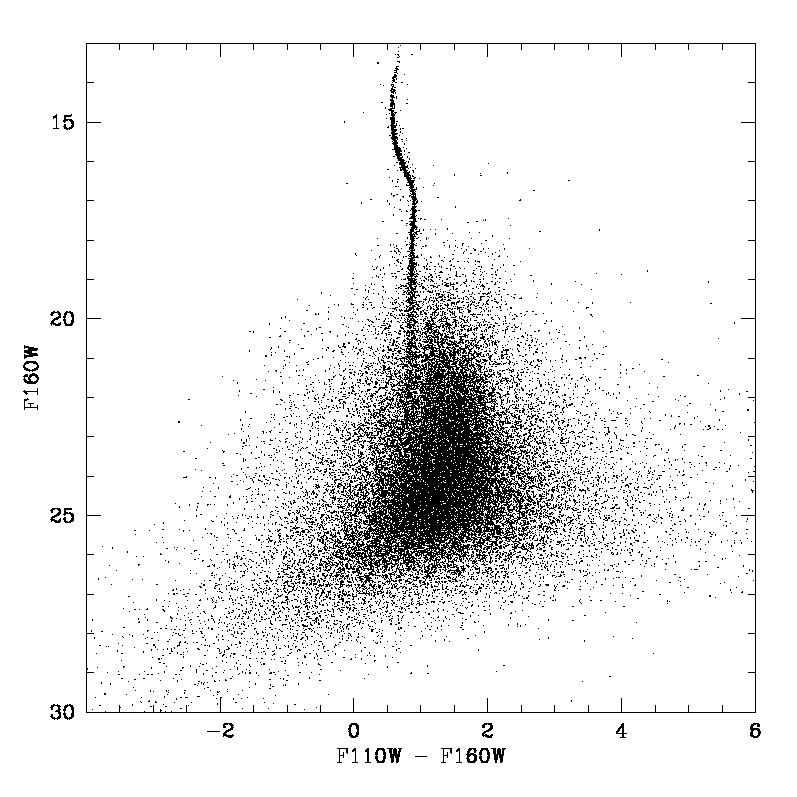}
}
\caption{NIR CMDs of all sources detected by {\tt DOLPHOT}. Left:
  $F110W$ vs. $F110W-F160W$, right: $F160W$ vs. $F110W-F160W$. Note
  that the CMDs are not selected and thus contain a large number of
  spurious detections like noise peaks and spikes around bright stars,
  but also faint sources that are not stars, and field stars that do
  not belong to the cluster. \label{cmdIRall}}
\end{figure*}

{\tt DOLPHOT} can run on data from multiple filters and cameras, thus
we were able to do the photometry on both $F110W$ and $F160W$ data
sets simultaneously.  The output file includes x and y positions,
photometric parameters like sharpness, crowding (the magnitude
difference to the measured magnitude if no neighbouring sources would
be fit simultaneously), the object type, and magnitudes in the VEGAmag
system for all sources detected in the $F110W$ and $F160W$ data.  A
calibrated NIR CMD is plotted in Fig.~\ref{cmdIRall}.  As can be seen,
the CMD is exceptionally deep. The total catalogue contained 51\,311
sources, but a large number of those will be spurious detections.

\subsection{Optical Data and photometry}
\label{optobs}

Among the archival {\it HST} material the only images which offer any hope of
detecting BDs in the optical are those deep and in the reddest filters
available, i.e., $F775W$ and $F814W$ (indeed, technically $>$2/3 of
these pass-bands are already in the NIR wavelength range).

We extensively searched in the {\it HST} archive, and identified four
$\sim$1200 seconds deep images taken with ACS/WFC in $F775W$ under
program GO-10146 (PI: Bedin) as the optimal for our purposes, as those
images were taken in low-sky mode, are well dithered, and collected in
a well defined epoch in 2005.48 (i.e., about 6.8 years before the
GO-12602 data).

Images were reduced with the software described in great detail in
Anderson et al.\ (2008).  Briefly, the method is essentially a
PSF-fitting where all pixels from all images are simultaneously fitted
using the appropriate PSFs which account for the spatial and temporal
shape-variation on each individual image. The key to the method is an
optimal knowledge on how to transform those pixels into a common
reference frame and an exquisite empirical PSF modeling. This software
is well tested and used in all the twelve works of the series ``The
ACS Survey of Galactic Globular Clusters'' (see, Sarajedini et al.\
2007).

The first photometric run returned 19990 optical detections. As the
detection of BDs might be only marginal in these optical images, we
relaxed the finding criteria in a second photometric run, imposing to
save all the local maxima, and not only the significant ones,
resulting in 1.5 million detections. Although most of these local
maxima likely are just fluctuations of the background noise, we can
still use them to set an optical upper limit to the NIR-detected BD
candidates.

For visual inspection of the WDs and BD candidates, stacked images
were created from the CTE-corrected FLT images. The stacked
images have pixels values resulting from median clipping of the
CTE-corrected FLT individual images, and are supersampled by a factor
of 2 in each direction (see Anderson et al.\ 2008 and Anderson \&
Bedin 2010 for further details).

\section{The Colour-Magnitude Diagrams}
\label{cmds}

We used the data in the two NIR filters $F110W$ and $F160W$ to create
a NIR CMD. As the full NIR catalogue contains a large number of
spurious detections, we created a best-photometry NIR CMD (described
below in Sect.~\ref{nircmds}). However, since WDs and BD candidates
have similar magnitudes and colours in the NIR, BD candidates cannot
be distinguished based on the NIR data alone. We then used additional
optical F775W data to create proper motion cleaned optical-NIR CMDs
(see Sect.~\ref{nircmds}) in which the WD and MS sequence are clearly
separate. Thus, the optical data are used to distinguish the WDs from
the BD candidates.

\subsection{Best-photometry NIR CMDs}
\label{nircmds}

\begin{figure*}
\centerline{
\includegraphics[width=\textwidth]{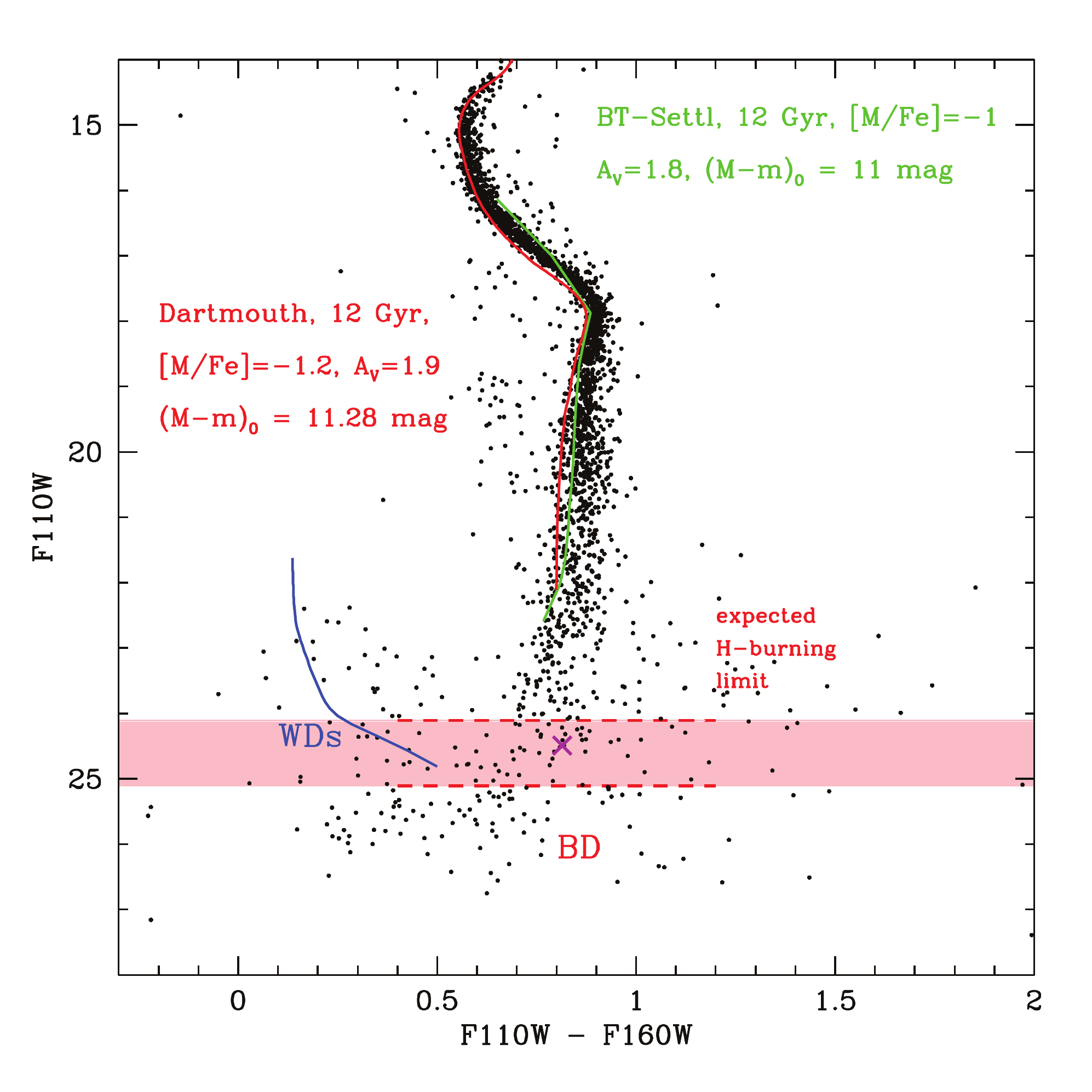}
}
\caption{NIR CMD that includes only best-photometry sources. The
  number of spurious detections should be minimized, however we expect
  that we did not include all true faint stellar sources that had
  originally been detected. The WD and BD regions have been labeled,
  and low-mass stellar models have been over-plotted. The expected end
  of the H-burning sequence is marked with red dashed lines and a
  shaded area. The magenta cross denotes the location of the known
  field BD SDSS-J125637.13-022452.4 \citep{burgasser2009}, scaled to
  M\,4's reddening and distance. Its position in the CMD supports that
  our data are indeed deep enough to reach well into the BD zone. See
  text for details. \label{cmdIRsel}}
\end{figure*}

In order to produce a clean CMD of only ``good'' stellar sources, we
selected the output catalogue in such a way that the faint sources
will be dominated by true stellar detections while keeping the number
of spurious detections at bay.{\bf \footnote{For this purpose, we
    settled on a selection that only includes source with object type
    2 or less (i.e., only ``stars''). We selected a sharpness between
    -0.05 and 0.05 (a sharpness of zero denotes a perfectly-fit star,
    a negative value indicates a broader source, a positive value
    indicates a source too ``sharp'', for example a cosmic ray - for
    an uncrowded field, sharpness values between -0.3 and 0.3 are
    recommended in the {\tt DOLPHOT} manual, so we apply a stricter
    selection criterion here). The crowding parameter indicates how
    much brighter a source would be if nearby stars would not have
    been measured simultaneously. A crowding of zero indicates an
    isolated star. We allowed for a crowding of no more than 0.3.}}
The resulting best-photometry catalogue includes 2526 sources, i.e,
$\approx 5\%$ of all detections. The best-photometry CMD is remarkably
clean, as can be seen in Fig.~\ref{cmdIRsel}. We can clearly see the
MS delineating down towards the expected H-burning limit and
going beyond that to fainter sources. Note that this CMD is not proper
motion cleaned, as we show all NIR sources that satisfy our selection
criterion on the photometry. All sources with magnitudes fainter than
24 mag in $F110W$ were visually inspected on the $F110W$ master image,
resulting in 177 visually confirmed NIR sources around or fainter than
the H-burning limit in $F110W$ (see Sect~\ref{Hmass} below).

The MS is narrow between the two MS ``knees'' ($F110W \approx 15$ mag
and $F110W \approx 18$ mag). The first ``knee'' occurs around
$T_{eff}\approx4500$ K and a mass of $\approx0.55 M_\odot$ and is due
to the formation of molecules in the cool stellar atmospheres, and the
second knee is a result of increasing electron degeneracy in the
(sub)-stellar interior close to the H-burning limit (Baraffe et
al. 1997). Below the second knee, the MS broadens towards fainter and
lower-mass MS stars. Indeed, the low-mass MS splits into two branches,
as shown in Milone et al.\ (2014), who had first pointed out multiple
stellar generations among VLMSs in M\,4, based on high-precision deep
optical {\it HST} data from the {\it HST} M\,4 core project and our
NIR data. The split can be clearly seen in the NIR CMD (see Milone et
al. 2014, their Fig. 2), as opposed to the optical data, demonstrating
that the NIR is also an ideal waveband to search for multiple
sequences along the lowest-mass and hence faintest MS. Note that the
goal in Milone et al.\ (2014) was to search for multiple generations
along the low mass MS based on very precise photometry. In contrast,
in this paper, our goal is to go deep and well beyond the H-burning
limit. 

In our Fig.~\ref{cmdIRsel}, we can see the MS delineating further down
towards the expected end of the H-burning sequence, marked with red
slashed lines and light-red shaded area. See Sect.~\ref{Hmass} below
for a discussion on the mass and NIR magnitude at the H-burning
limit. Around $F110W \approx 24.5$ mag the number of sources on the MS
decreases and the MS peters out.  On the blue side of the faint MS,
the WD sequence can be seen, starting around $F110W < 22$ mag and
$F110W - F160W \approx 0.2$ mag and going fainter and redder.

An increase in source number seems to be apparent below $F110W > 25$
mag, i.e., below the expected end of the H-burning sequence. This is
the area in the CMD where we expect the BDs to appear. This happens to
coincide with the WD sequence as well, i.e., this is the region where
WD and BD cooling sequences would be expected to cross. Unfortunately,
this also means that we cannot disentangle WDs and BD candidates based
on our NIR data alone. In order to help with both cluster membership
determination and distinguishing WDs and BD candidates, the deep
optical data from GO--10146 were used (see Sect.~\ref{optobs} and
Sect,~\ref{optnircmds}).

The WD as well as the BD regions have been indicated in
Fig.~\ref{cmdIRsel}. For orientation purposes, we have plotted a WD
sequence (blue line). The WD cooling sequence was constructed by
interpolating on the Wood (1995) grid of theoretical WD cooling
curves, adopting a mean WD mass of $0.55 M_\odot$. Using a grid of
synthetic DA WD spectra kindly provided by B. G\"ansicke (see
G\"ansicke et al.\ 1995) we carried out synthetic photometry with {\tt
  PySynphot}. Note that we have shifted the WD cooling sequence to get
a reasonable match to the underlying CMD. This required a rather large
distance of 2.2 kpc and a reddening of $E(B-V)=0.55$ mag (a standard
reddening law (Seaton 1979) is built into {\tt PySynphot}). Our
cooling sequence starts at $T_{eff} = 50\,000$ K and terminates at
$T_{eff} = 8\,000$, but note that the coolest WDs in M\,4 have
temperatures as low as $T_{eff} = 4\,000$ (Bedin et al. 2009).

In addition, we have marked the location of the known field BD
SDSS-J125637.13-022452.4 \citep{burgasser2009}, which has a
metallicity similar to M\,4 but is likely several Gyr younger. As a
consequence, its cooling time is shorter and thus it is expected to be
brighter than the M\,4 BDs. The observed $J$- and $H$-band magnitudes
agree with a 5 Gyr old, $0.078 M_\odot$ source at a metallicity of
[M/H]=-0.5 dex. We scaled the observed $J$- and $H$-band magnitudes to
the WFC3 NIR filters and applied M\,4's distance and reddening
(following Hendricks et al.\ 2012, we adopted a reddening of $E(B-V) =
0.37$ mag, a true distance modulus of 11.28 mag, $R = 3.67$, $A_J =
0.302 * A_V$ and $A_H = 0.191 * A_V$) and over-plotted the field BD on
our CMD. Its location supports that our data are indeed deep enough to
reach well into the BD zone.

Two 12 Gyr isochrones have been overplotted on the best-photometry
CMD, a BT-Settl model based on the Asplund et al.\ (2009) solar
abundances and the Barber \& Tennyson (2006) line list (Allard et
al. 2012), for a metallicity of $\rm{[M/H]}=-1$ dex; and a Dartmouth
model for $\rm{[Fe/H]}=-1.2$ and $\rm{[\alpha/Fe]}=+0.4$ (Dotter et
al.\ 2008). Note that we do not attempt to derive cluster parameters
from the isochrone fitting, instead, we have fit the isochrones by eye
so that they best overlap with the underlying CMD\footnote{The best
  fit was achieved with the reddening law from Hendricks et al. 2012
  but a larger $A_V = 1.9$ for the Dartmouth isochrone. For the
  BT-Settl isochrone, we chose a smaller distance modulus of 11 mag
  and $A_V = 1.8$. The difference in shape of the isochrones, as well
  as the difference in the best-fit parameters, reflect the
  differences in the underlying physics, i.e. treatment of the stellar
  atmospheres including molecules. For a more in-depth discussion of
  the input physics to the models we refer the reader to the BT-Settl
  and Dartmouth webpages and the references given there
  (https://phoenix.ens-lyon.fr/Grids/BT-Settl/ and
  http://stellar.dartmouth.edu/models/).}. Note also that both sets
terminate at a stellar mass of 0.083 $M_\odot$ (BT-Settl) or 0.1
$M_\odot$ (Dartmouth), i.e., they do not reach to the H-burning limit.

\subsection{Optical-NIR CMDs}
\label{optnircmds}

\begin{figure*}[h]
\includegraphics[width=\textwidth]{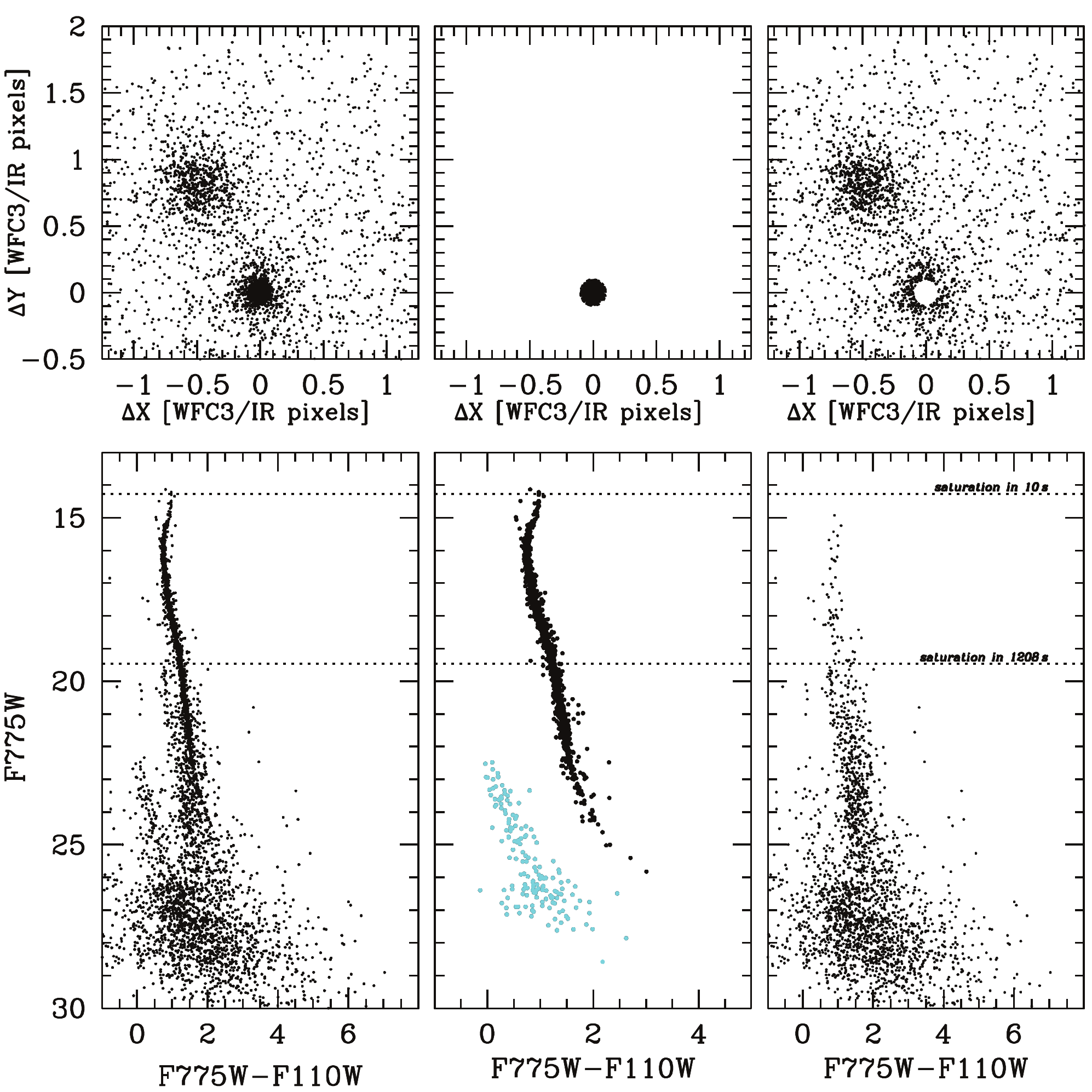}
\caption{Top row: Vector point diagrams for all sources with
  counterparts with a displacement of no more than 2 WFC3/IR
  pixels. Bottom row: Optical-NIR ($F775W-F110W$) CMDs for all sources
  with optical counterparts (left); only sources with a displacement
  of less than 0.1 pixels, suggesting that they are cluster members
  (middle); and for the remaining field stars (right). The cluster CMD
  is used to select WDs, plotted in light blue. \label{cmdj}}
\end{figure*}

\begin{figure*}[h]
\includegraphics[width=\textwidth]{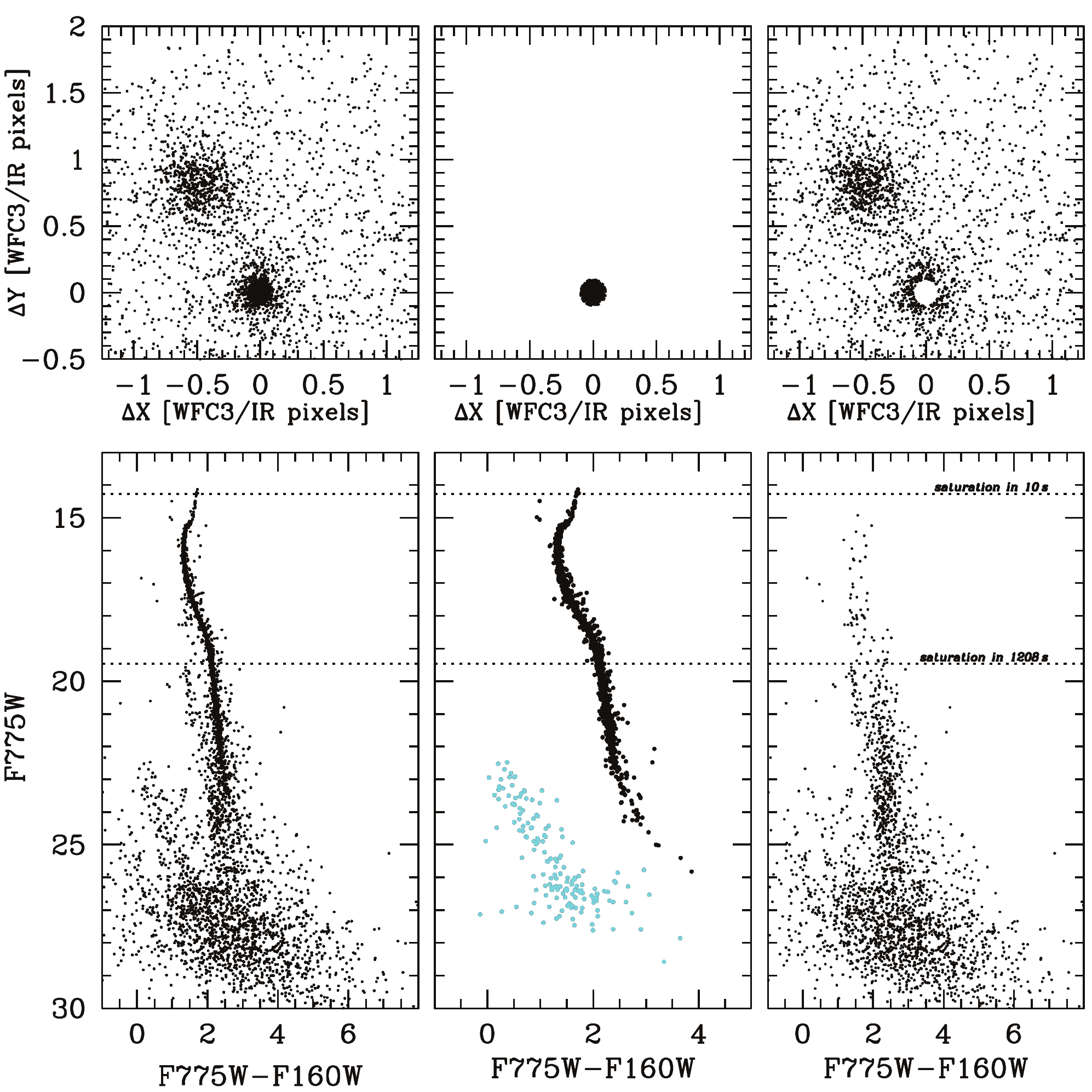}
\caption{The same as in Fig.~\ref{cmdj}, but for $F775W-F160W$. \label{cmdh}}
\end{figure*}

Our NIR and optical catalogues have been matched using a six-parameter
linear transformations between the star positions in the different
epochs. As reference stars for the transformations we used only
well-measured, isolated, non-saturated cluster stars with a high
signal-to-noise and low residuals. The {\it predicted positions} of
the first epoch sources in the second epoch are compared with the {\it
  observed positions} and the displacements between first and second
epoch are calculated. The top panel in Figures~\ref{cmdj}, \ref{cmdh},
\ref{pmcmdIR110} and \ref{pmcmdIR160} shows the displacements in
WFC3/IR, based on the total {\tt DOLPHOT} NIR (51\,311 detections)
and optical (19990 sources) catalogue. Since cluster stars have been
used for the reference list, we expect cluster members to agglomerate
around zero in the displacement vector point diagram. Indeed, two
populations can be distinguished: a dense and tight agglomeration of
data points around $\Delta Y = 0$ and $\Delta X = 0$ which denotes the
cluster members, and a more widely spread data region centering around
$\Delta Y \approx 0.75$ and $\Delta X \approx -0.5$ which denotes
field sources. The latter are mostly Bulge sources, reflecting the low
tangential motion of M\,4 around the Galactic center (see also Bedin
et al.\ 2003).

The corresponding CMDs are plotted in the second row in
Figures~\ref{cmdj}, \ref{cmdh}, \ref{pmcmdIR110} and
\ref{pmcmdIR160}. The left CMDs show all sources within up to two WFC3
pixels displacement, the middle CMDs show only sources with a
displacement of no more than 0.1 pixels which suggests that they are
cluster members, and all non-cluster sources are shown in the right
diagrams. A displacement of 0.1 pixels corresponds to a proper motion
of 1.9 mas/year, based on our timeline of 6.8 years and a pixel scale
of 0.13$\arcsec$. This agrees well with e.g.\ Bedin et al.\ (2003,
2009) and Zloczewski et al.\ (2012). As can be seen, the proper-motion
cleaned ``cluster'' CMDs (middle diagrams) show a well defined MS,
terminating around $F775W\approx26$ mag in both CMDs in
Figures~\ref{cmdj} and ~\ref{cmdh}, as well as a well defined WD
sequence (light-blue data points) going down to the bottom of the WD
sequence around $F775W\approx28$ mag.

On the other hand, the proper-motion cleaned optical-NIR CMDs do not
show any BD candidates, i.e., sources fainter than $F775W > 28$ mag
and redder than $F775W - F110W > 3.5$ or $F775W - F160W > 4.5$ mag
(see also Fig.~\ref{models}). This was expected, as the deep optical
CMD presented in Bedin et al.\ (2009) did not show any potential MS
sources fainter than the H-burning limit. However, and most
importantly in the context of this paper, the BDs are expected to be
much redder than the WDs in the optical-NIR CMDs. Indeed, the WD
sequence and the MS are clearly separated in our optical-NIR
CMDs. Therefore we can use the deep optical data set to identify the
WDs in our NIR CMDs and disentangle WDs from potential BDs.

\subsection{The minimum mass at the Hydrogen-burning limit}
\label{Hmass}

What is the minimum mass at the Hydrogen-burning limit in a
low-metallicity cluster like M\,4? And, as a consequence, at which NIR
magnitude and colour do we expect the H-burning limit in our NIR CMD?
Early theoretical work (Kumar 1963) suggested a lower limit to the
stellar MS at $\approx 0.07 M_\odot$ for population I and $\approx
0.09 M_\odot$ for the metal-poor population II stars, to which also
GCs belong. Hayashi \& Nakano (1963) suggested that stars less massive
than $0.08 M_\odot$ cannot undergo Hydrogen burning. They further
suggested that the limiting mass is not very different for Population I
and Population II stars. Burrows et al.\ (1993) presented a zero
metallicity theoretical model (i.e., Population III) which suggests a
limiting H-burning mass as high as $0.094 M_\odot$. Treatment of the
atmosphere has a considerable impact on the predicted limiting mass,
as can be seen in Chabrier et al.\ (2000) who suggested a limiting
mass of $0.065 M_\odot$ for models that include dust formation, Saumon
\& Marley (2008) who suggested a limiting mass of $0.075 M_\odot$ for
cloudless models, and $0.070 M_\odot$ for cloudy models; all for solar
metallicity and an age of 10 Gyr.

Previous theoretical works suggest that the H-burning limit is at
higher masses for more metal-poor stars. The models roughly agree on a
limiting mass of $0.075 M_\odot$ for solar metallicities. Following
Hayashi \& Nakano (1963), we conservatively assume that the H-burning
mass for a population as metal-poor as M\,4 is between $0.075 M_\odot$
and $0.08 M_\odot$. 

Unfortunately, detailed sub-stellar models for sub-solar metallicities
are presently not available. Updated models that extend well into the
BD regime are currently being computed (Allard, private communication). 

\begin{figure*}
\centerline{
\includegraphics[width=\textwidth]{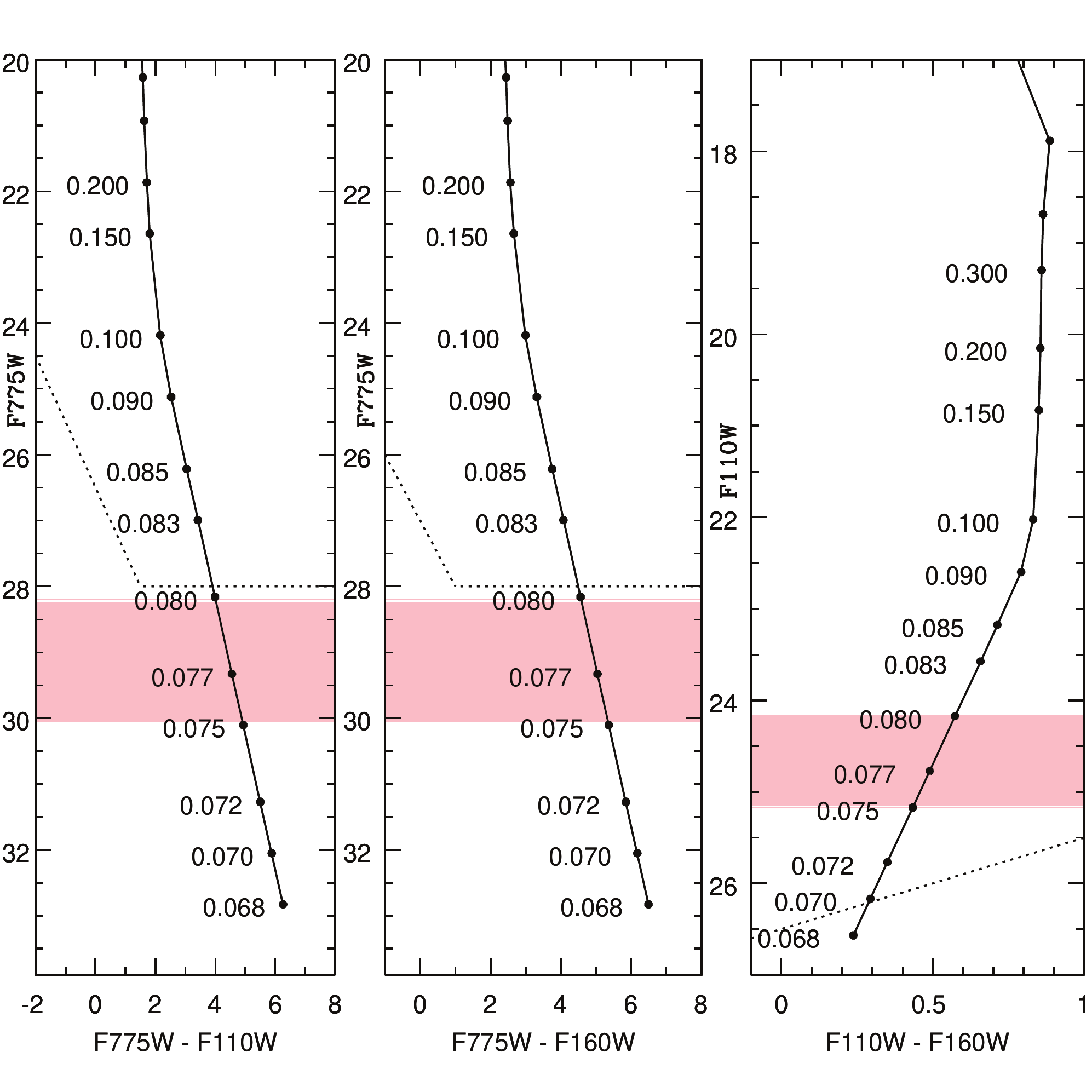}
}
\caption{12 Gyr BT-Settl models (Allard et al.\ 2012, 2013) in {\it
    HST}/WFC3 filters, scaled to M\,4's distance, reddening and
  metallicity. The models are extrapolated into the sub-stellar regime
  to 0.068 $M_{\odot}$. Masses are indicated along the sequences. The
  end of the H-burning sequence is estimated between 0.075 and 0.08
  $M_{\odot}$ and is indicated with a light-red shaded
  area. Detection limits in our NIR and optical data are
  indicated with dotted lines. See text for further details.\label{models}}
\end{figure*}

\begin{figure*}
\centerline{
\includegraphics[width=\textwidth]{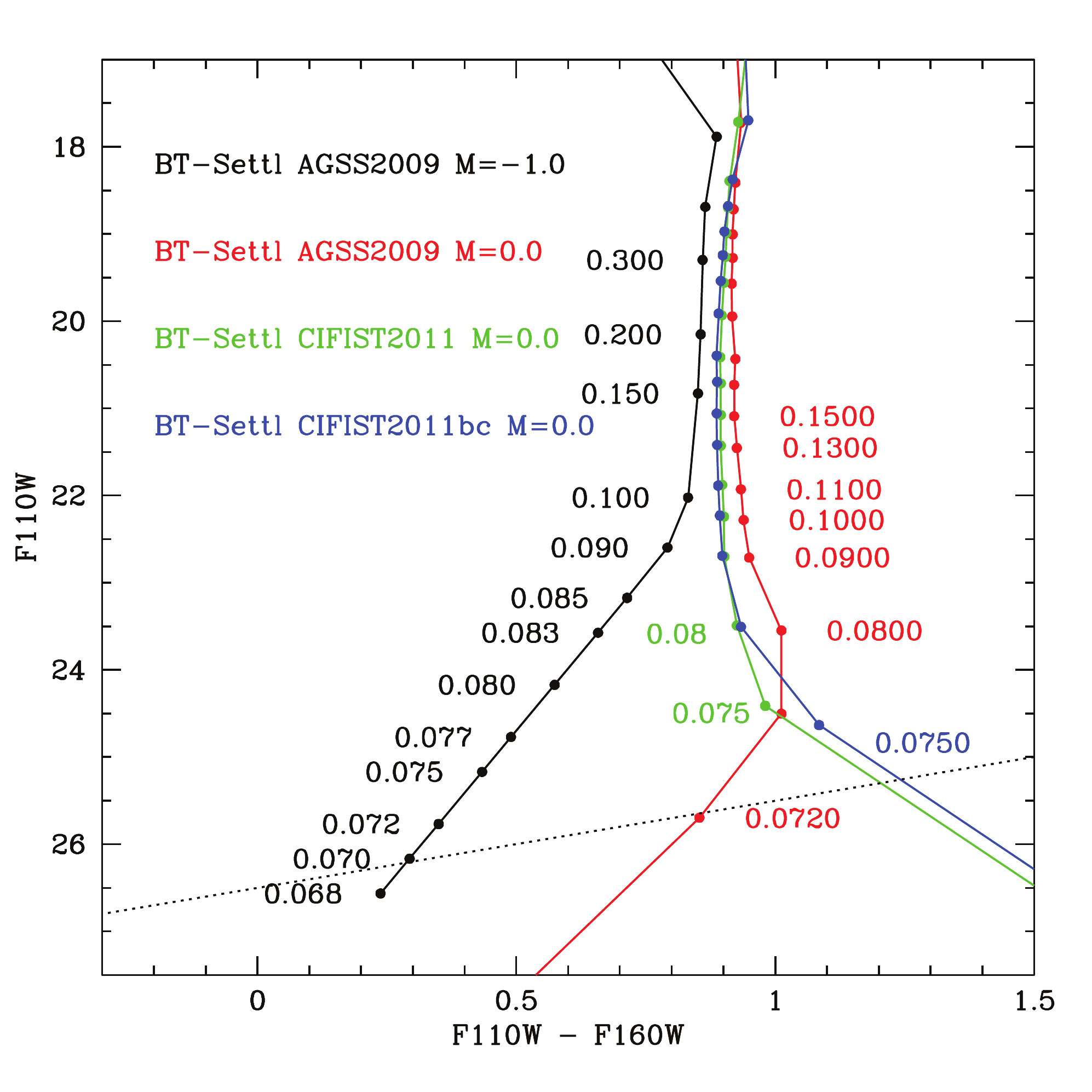}
}
\caption{Comparison of our extrapolated 12 Gyr BT-Settl model with
  sub-stellar models at a metallicity of $\rm{[M/H]} = 0.0$ dex, and
  with different solar abundances (AGSS2009 refers to the Asplund et
  al.\ 2009 solar abundance, CIFIST2011 was based on Caffau et al.\
  2011, CIFIST2011bc includes additional adjustments. See the PHOENIX
  webpage at {\tt https://phoenix.ens-lyon.fr/Grids/BT-Settl} for more
  information). The different input physics are reflected in the
  different shapes of the models, however, the expected end of the
  H-burning sequence between $0.075 M_\odot$ and a $0.08 M_\odot$ is
  around $F110W \approx 24$ mag for all models. \label{compare}}
\end{figure*}

However, the most recent set of BT-Settl models (Allard et al.\ 2012;
2013) suggest a strong metallicity dependence of the shape and
luminosity of the low-mass MS. These models only go down to $0.083
M_\odot$ and are close to the H-burning limit (Allard private
communication), but do not go beyond the stellar sequence into the
sub-stellar regime. Thus, we linearly extrapolated the BT-Settl models
(based on the Asplund et al.\ 2009 solar abundances) for $\rm[M/H]=-1$
dex and an age of 12 Gyr (closest to M\,4's parameters) down to
sub-stellar masses of $0.068 M_\odot$, and applied distance and
reddening as in Fig.~\ref{cmdIRsel} for the BT-Settl model. The
extrapolated models are plotted in Fig.~\ref{models}, and the
magnitudes around the H-burning limit are listed in
Table~\ref{mlim}. As mentioned above, no sub-stellar models for low
metallicities are currently available. Different models exist for a
metallicity of $\rm{[M/H]} = 0.0$ dex. Thus, in Fig.~\ref{compare} we
compare our extrapolated models with these more metal-rich models,
all for an age of 12 Gyr and scaled to M\,4's distance and
reddening. The effect of the different atmospheric physics can be
clearly seen in the shape and colour of the models. However, the
expected end of the H-burning sequence between a $0.075 M_\odot$ and a
$0.08 M_\odot$ is at comparable magnitudes in all sub-stellar
models. This gives us some confidence that we can use the extrapolated
metal-poor BT-Settl models to get an estimate of the magnitude and
colour range of the H-burning limit. For more exact values we will
have to wait for the updated metal-poor models, but we remind the
reader that the main purpose of this project is to provide the
metal-poor benchmark sources and thereby fill the observational plane
with data points that are needed to constrain theoretical models.

\begin{figure*}
\centerline{
\includegraphics[width=7.5cm, height=8cm]{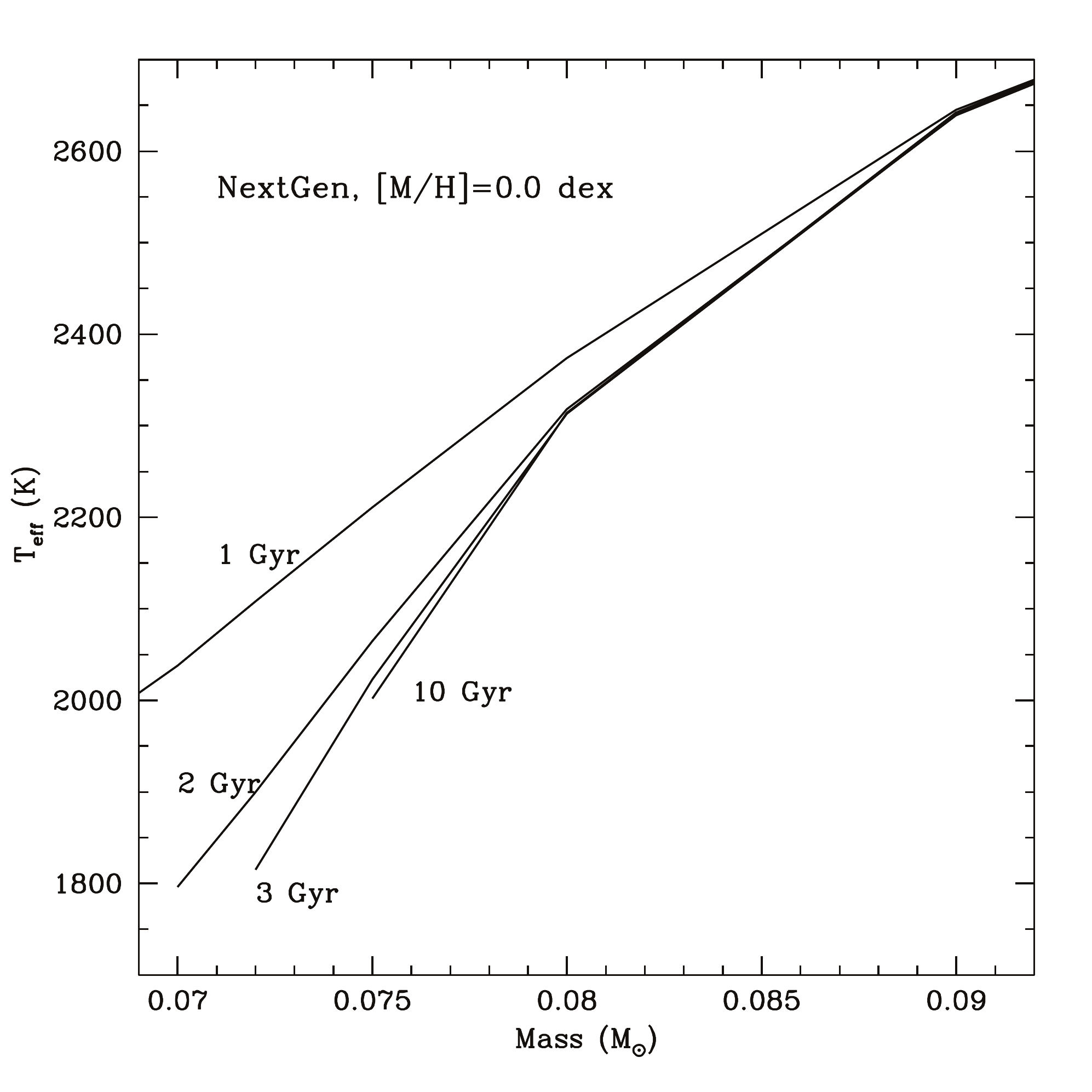}
\includegraphics[width=7.5cm, height=8cm]{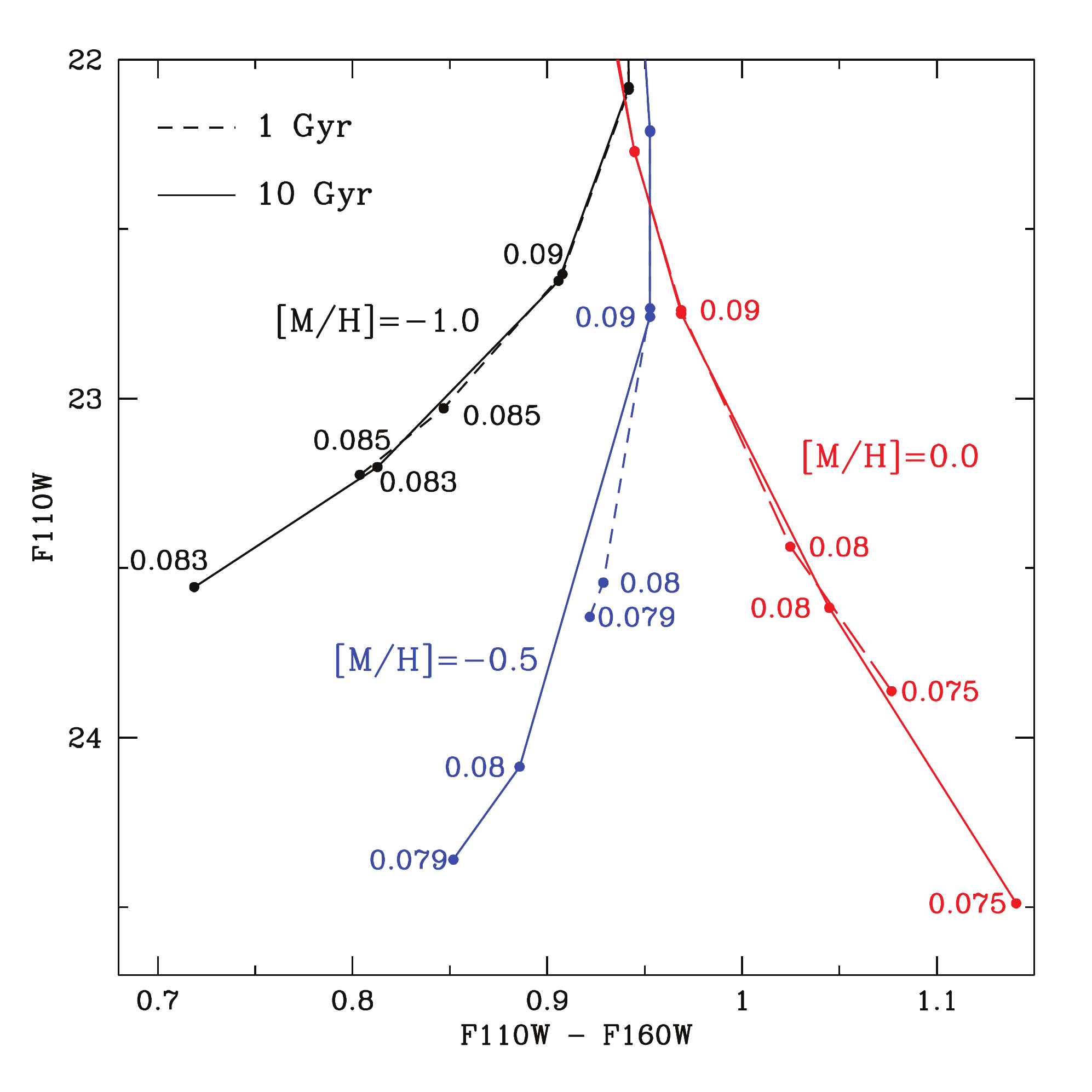}
}
\caption{Low-mass stellar/BD models (NextGen) based on Baraffe et al.\
  (1997, 1998), and Hausschildt et al. 1999. Left: Mass vs. $T_{eff}$
  for 1 Gyr, 2 Gyr, 3 Gyr and 10 Gyr ([M/H]=0.0 dex). Since BDs cannot
  sustain their luminosities via nuclear processes, they cool with
  time and become fainter. Right: Low-mass isochrones that show the
  effect of BD cooling. The dashed line is a 1 Gyr isochrone, the
  solid line a 10 Gyr isochrone. Substellar models at low
  metallicities currently do not exist, thus isochrones for different
  metallicities are compared: [M/H]=-1 dex isochrones plotted in
  black, [M/H]=-0.5 dex in blue, [M/H]=0 dex in red. The [M/H]=-1 dex
  10 Gyr isochrone terminates at 0.083 $\rm{M}_\odot$, suggesting that
  such a low-mass object already cooled and faded by 0.3 mag in
  $F110W$. Lower masses are available for higher metallicities,
  suggesting that a 0.079 $\rm{M}_\odot$ source fades by 0.7 mag at a
  metallicity of [M/H]=-0.5 dex, about the same as a 0.075
  $\rm{M}_\odot$ source at a metallicity of [M/H]=0 dex. The models
  suggest that lower-mass sources cool faster the lower the mass, and
  also the lower the metallicity. \label{BDcool}}
\end{figure*}

Unlike stars, BDs cannot retain their luminosities via nuclear
fusion. As a consequence, they cool with time, and a BD will be at
fainter luminosities in an old GC compared to a young BD of the same
mass and metallicity. To get an estimate of this effect, we used
NextGen models (Baraffe et al. 1997, Baraffe et al. 1998; see
Fig.~\ref{BDcool}). In the left diagram, we plot mass against
$T_{eff}$ for various ages. The right diagram shows 1 Gyr and 10 Gyr
NextGen models for different metallicities. Since substellar
isochrones at M\,4's low metallicity of [M/H]=-1 dex and ages of $\geq
10$ Gyr currently do not exist, we also plot isochrones for a
metallicity of [M/H]=-0.5 and 0.0 dex which go down to lower
masses. As can be seen, metallicity has a considerable impact on the
shape of the isochrones as well as on the cooling time scale and hence
fading of low-mass, substellar objects. The [M/H]=0 dex isochrones
continue to extend with time to fainter magnitudes and redder colours,
suggesting that a 0.075 $\rm{M}_\odot$ source fades by $\approx 0.7$
mag in $F110W$ from 1 to 10 Gyr, and a 0.08 $\rm{M}_\odot$ source
becomes fainter by $\approx 0.2$ mag. The [M/H]=-0.5 dex isochrones
suggest that a source of 0.079 $\rm{M}_\odot$ becomes fainter by
$\approx 0.7$ mag, and a 0.08 $\rm{M}_\odot$ source becomes fainter by
$\approx 0.5$ mag, but the sources also become bluer rather than
redder. The [M/H]=-1 dex 10 Gyr isochrones terminates at 0.083
$\rm{M}_\odot$. According to the models, such a metal-poor, low-mass
source already fades by 0.3 mag in $F110W$ from 1 to 10 Gyr. The
models suggest that the blue-turn is more pronounced for
lower-metallicities, and also metal-poor sources become fainter
compared to metal-richer sources at the same mass, i.e. they cool
faster.

Our optical-NIR CMDs presented in Figures~\ref{cmdj} and \ref{cmdh} show
that the observed MS peters out around $F775W\approx26$ mag,
suggesting that we are approaching the end of the H-burning
sequence. Our NIR CMD in Fig.~\ref{cmdIRsel} suggests that the MS
peters out just below $F110W\approx24$ mag where the density of stars
decreases. At fainter magnitudes, the WD sequence crosses the MS and
the star density increases again. Based on the CMDs, we estimate our
detection limits around $F775W\approx28$ mag, $F110W\approx26.5$ mag
and $F160W\approx27$ mag. The detection limits are also indicated in
Fig.~\ref{models} with a dotted line. Comparing the theoretical models in
Fig.~\ref{models} with our observed CMDs and taking the detection
limits and the predicted H-burning limit into account, we conclude that
we have reached beyond the H-burning limit in our NIR CMD and are
probably just above or around this limit in our optical-NIR CMDs.

\begin{table*}
\begin{center}
  \caption{\label{mlim} Masses and corresponding magnitudes around the
    H-burning limit, estimated from the extrapolated BT-Settl models
    and scaled to M\,4's distance and reddening. All magnitudes are
    for {\it HST}/WFC3 filters and in the VEGAmag system.}
\begin{tabular}{cccc}
\tableline
mass [$M_\odot$] & $F110W$ [mag] & $F160W$ [mag] & $F775W$ [mag]\\  
\tableline\tableline    
0.075           & 25.170       & 24.736     & 30.105\\ 
0.077           & 24.771       & 24.281     & 29.327\\ 
0.080           & 24.172       & 23.598     & 28.160\\ 
\tableline
\end{tabular}
\end{center}
\end{table*}

\subsection{BD candidates}
\label{bdcands}

As with the best-photometry faint NIR sources, we visually inspected
all cluster WD candidates, i.e., sources whose proper motion (or
rather displacements) suggest that they are cluster members and whose
position in the optical-NIR CMDs suggest that they are WDs. Our first
photometric run on the optical data did not return an optical
counterpart for 59 of the best-photometry faint NIR sources. We then
over-plotted the positions of those 59 faint NIR sources (i.e., {\it
  without} an initial optical counterpart) on the optical $F775W$
master image and inspected each position by eye.  Visual inspection of
the $F775W$ master image showed a faint optical source at or close to
the location of the NIR source in most cases. Thus, for the second
optical photometric run (see Sect.~\ref{optobs}), the parameter
settings were relaxed so that all local maxima were retained,
resulting in 1.5 million detections. Nearly all of those are just
spurious detections (i.e., background noise or spikes in PSF wings
that are not real stellar sources), however, a further 47 optical
counterparts to the faint best-photometry NIR sources were detected
and thus were added to the initial list of optical-NIR matches.

For the remaining twelve faint NIR sources, still no optical
counterpart was returned. However, out of those, five are located on
PSF streaks and two in the ACS WFC chip gap, so that nothing can be
said about an optical counterpart. For four of the remaining five
sources, visual inspection of the $F775W$ master image seem to
indicate an optical source on the position of the NIR source, but no
photometric measurement was possible. However, we used nearby stars of
similar brightness (based on pixel counts) to estimate magnitude
limits. One of these sources appears to have an optical counterpart in
the centre of our search circle, probably at $F775W\approx26$
mag. This is too bright for a BD candidate and thus makes this source
a WD candidate. Thus, we do not consider it further. The remaining
four sources, id 1 to 4, with no optical photometry are listed in
Table~\ref{bdcand}, and images are shown in Fig.~\ref{bdimas}. For
comparison, we also show images of WDs selected from the proper motion
cleaned CMD and which have similar NIR magnitudes to the four NIR
sources without optical photometry (see Fig.~\ref{wdimas} and
Table~\ref{wdcand}).

At the rim of our search circle for source 1, an optical source can be
seen. This, however, is too far away ($0\farcs1$ away from the
position of the NIR source) and would not agree with being a cluster
member. In the centre of our search circle, an optical source just
might be visible. If true, this source would have $F775W\gtrsim28$
mag, i.e., the local detection limit. For source 2, an optical
counterpart is visible in the centre of the search circle, again this
source would be close to the detection limit at $F775W\gtrsim28$
mag. Thus, both Source 1 and 2 are likely (massive) BD
candidates. Unfortunately, Source 2 is close to a saturation streak
from a nearby bright stars in the optical image. The potential
counterpart, however, can be clearly distinguished.  Source 3 shows a
``bright'' optical source at the rim of our search circle, probably at
$F775W\approx25.8$ mag, again based on the magnitudes of nearby stars
that appear to be of similar brightness. However, this optical source
is $0\farcs15$ away from the position of the NIR source and thus too
far away to agree with being a cluster member. A very faint optical
source might just be in the centre of the search circle, if true, then
this optical source would be at $F775W>28.6$ mag, the local detection
limit, making source 3 a good BD candidate. Just one source, source 4,
does not show an optical counterpart at the location of the NIR source
(i.e., within our search circle), and is thus our best BD
candidate. Since the optical photometry did not return magnitudes
fainter than $F775W = 32.9$ mag, an optical counterpart to BD
candidate 4 must be fainter than this absolute optical limit. Note
that Source 4 appears somewhat extended and could probably consist of
two or three faint sources, or possibly an extended object (although
we expect galaxies to be much bluer, see Bedin et al.\ 2009).

\begin{figure*}[h]
\includegraphics[width=\textwidth]{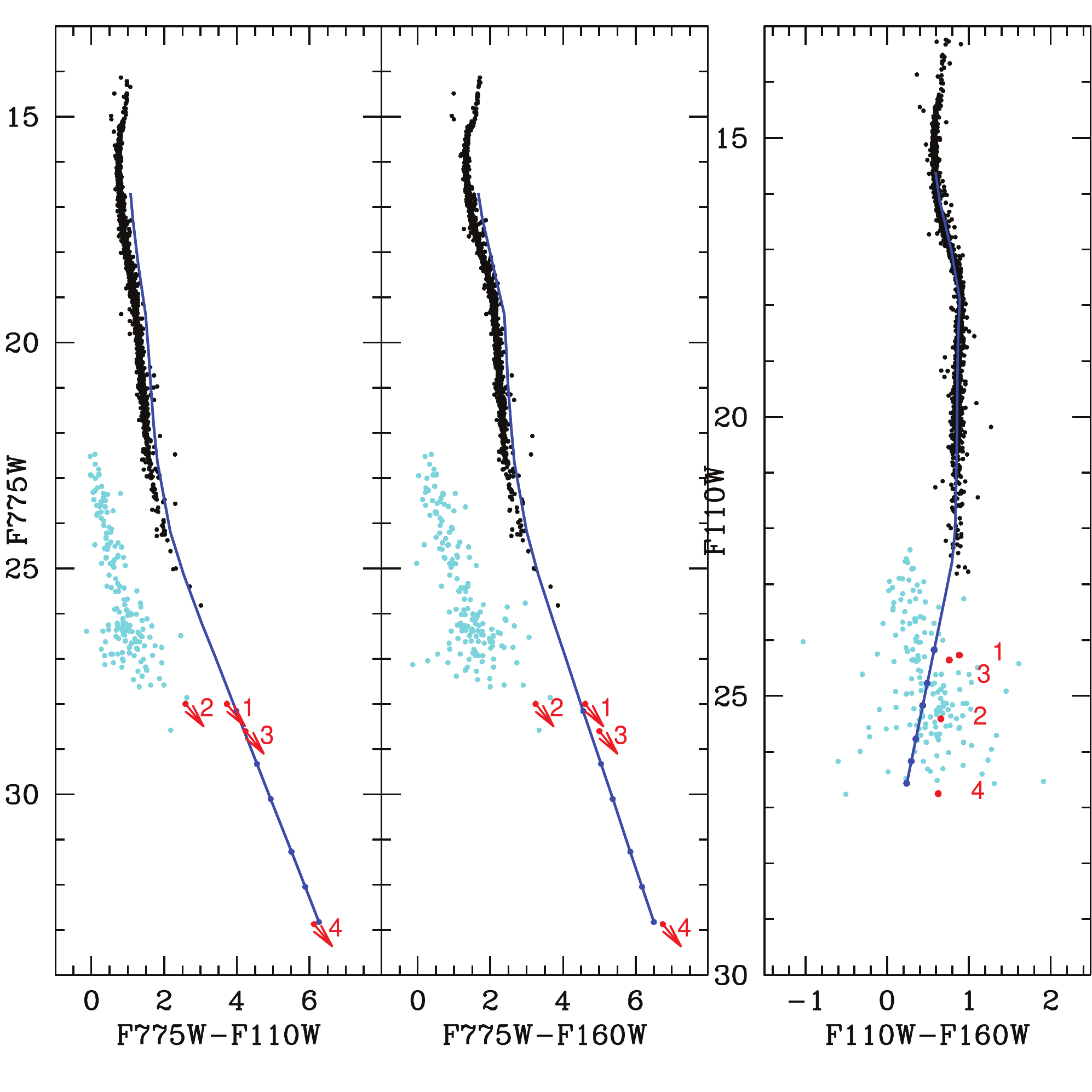}
\caption{Optical-NIR CMDs (left and middle) and NIR CMD of all sources
  with an optical counterpart. We also show the position of the four
  BD candidates, marked with red arrows, assuming the upper optical
  magnitude limits given in Table~\ref{bdcand}. We also overplot the
  extrapolated BT-Settl 12 Gyr isochrone. As can be seen, the position
  of all four BD candidates is very close to the isochrone, suggesting
  that they are indeed BDs. See the text for more
  details. \label{pmcmds}}
\end{figure*}

The positions of the four BD candidates are indicated in the CMDs in
Fig.~\ref{pmcmds}. Since we do not have an optical measurement, we
provide the upper optical magnitude limits given in
Table~\ref{bdcand}. We also overplot the extrapolated BT-Settl 12 Gyr
isochrone, using the distance and reddening parameters derived from
the best fit in the NIR CMD. As can be seen, all four BD candidates
are very close to the extension of the MS into the BD regime, which
supports our classification of these sources as good BD
candidates. Source 2 is blueward of the isochrone. Its position in the
optical-NIR CMDs agrees with this source being a faint WD at the very
bottom of the (optical) WD cooling sequence in M\,4, but it also
agrees with this source being a VLMS star or a massive BD. In the NIR
CMD this source is {\it not} at the very bottom of the WD sequence. We
suggest that this source is a good BD candidate. Sources 1, 3 and 4
all are very close to the MS, and their position in the optical-NIR
CMDs does not suggest that these sources are WDs, but rather BDs. WDs
are much bluer and brighter than our BD candidates.

\begin{table*}
\begin{center}
  \caption{\label{bdcand} Best-photometry NIR sources fainter than
    $F110W > 24$ mag, and for which the optical photometry did not
    return a counterpart. The magnitudes and errors (cols. 4 and 5)
    are the weighted mean and error of the weighted mean as returned
    from {\tt DOLPHOT}. See text for further details.}
\begin{tabular}{lllllll}
\tableline
id$_{BD}$ & RA           & DEC   & $F110W$        & $F160W$        & $F775W$     & comment\\
 & [h:m:s] & [$^\circ:\arcmin:\arcsec$] & [mag]   & [mag]          & [mag]       & \\
\tableline\tableline
1 & 16:23:41.701 & -26:29:17.82 & $24.27\pm0.02$ & $23.39\pm0.11$  & $\gtrsim$28 & BD candidate\\ 
2 & 16:23:45.443 & -26:30:05.58 & $25.41\pm0.05$ & $24.75\pm0.50$  & $\gtrsim$28 & WD/BD candidate\\ 
3 & 16:23:44.711 & -26:29:30.90 & $24.36\pm0.02$ & $23.60\pm0.11$  & $>28.6$     & BD candidate\\ 
4 & 16:23:45.371 & -26:29:37.40 & $26.75\pm0.16$ & $26.13\pm0.18$  & $>$32.9     & BD candidate\\ 
\tableline
\end{tabular}
\end{center}
\end{table*}

\begin{table*}
\begin{center}
  \caption{\label{wdcand} Same as Table~\ref{bdcand} but for WDs
    selected from the optical-NIR CMDs and with similar NIR magnitudes
    than the BD candidates. Their optical counterpart, however, is
    clearly detectable. The magnitudes and errors (cols. 4 to 6) are
    the weighted mean and error of the weighted mean as returned from
    {\tt DOLPHOT} for the NIR, and the median-clipped mean and
    standard deviation for the optical magnitudes.}
\begin{tabular}{llllll}
\tableline
id$_{WD}$ & RA   & DEC           & $F110W$       & $F160W$         & $F775W$ \\
  & [h:m:s]      & [$^\circ:\arcmin:\arcsec$] & [mag] & [mag]      & [mag] \\
\tableline\tableline
1 & 16:23:44.354 & -26:30:26.09 & $25.70\pm0.06$ & $25.10\pm0.24$ & $27.184\pm0.44$ \\ 
2 & 16:23:36.848 & -26:29:33.58 & $26.30\pm0.11$ & $25.62\pm0.40$ & $27.618\pm0.44$ \\ 
3 & 16:23:36.953 & -26:29:33.81 & $25.67\pm0.06$ & $25.14\pm0.48$ & $26.607\pm0.23$ \\ 
4 & 16:23:47.062 & -26:30:37.80 & $25.68\pm0.07$ & $25.04\pm0.37$ & $26.368\pm0.90$ \\ 
\tableline
\end{tabular}
\end{center}
\end{table*}

\begin{figure*}[h]
\includegraphics[width=\textwidth]{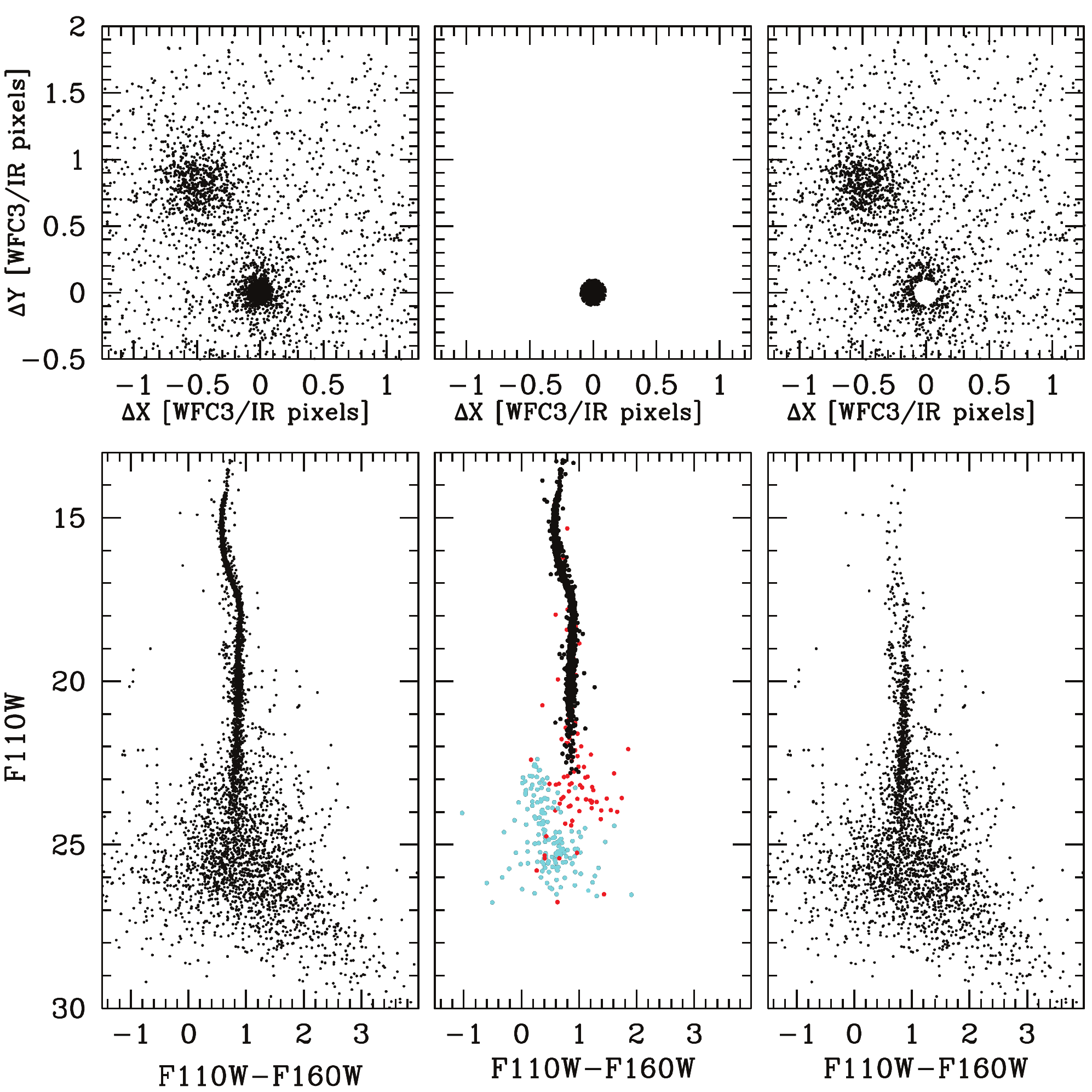}
\caption{Top row: Vector point diagram for all sources with
  counterparts with a displacement of no more than 2 WFC3/IR
  pixels. Bottom row: NIR CMDS ($F110W$ vs. $F110W-F160W$) CMDs for
  all sources with optical counterparts (left); only sources with a
  displacement of less than 0.1 pixels, suggesting that they are
  cluster members (middle); and for the remaining field stars
  (right). The WDs selected from the optical-NIR CMDs in Figures
  \ref{cmdj} and \ref{cmdh} are plotted in light-blue, MS stars in
  black. Best-photometry NIR sources {\it without} an optical
  counterpart are plotted in red. As can be seen, the NIR
  best-photometry contains many VLMSs down to the expected H-burning
  limit, and reveals twelve sources around and below the expected
  H-burning limit ($F110W\geq24$ mag). Out of these, one is a likely WD
  candidate, whereas four sources are good BD candidates. See the text
  for further details. \label{pmcmdIR110}}
\end{figure*}

\begin{figure*}[h]
\includegraphics[width=\textwidth]{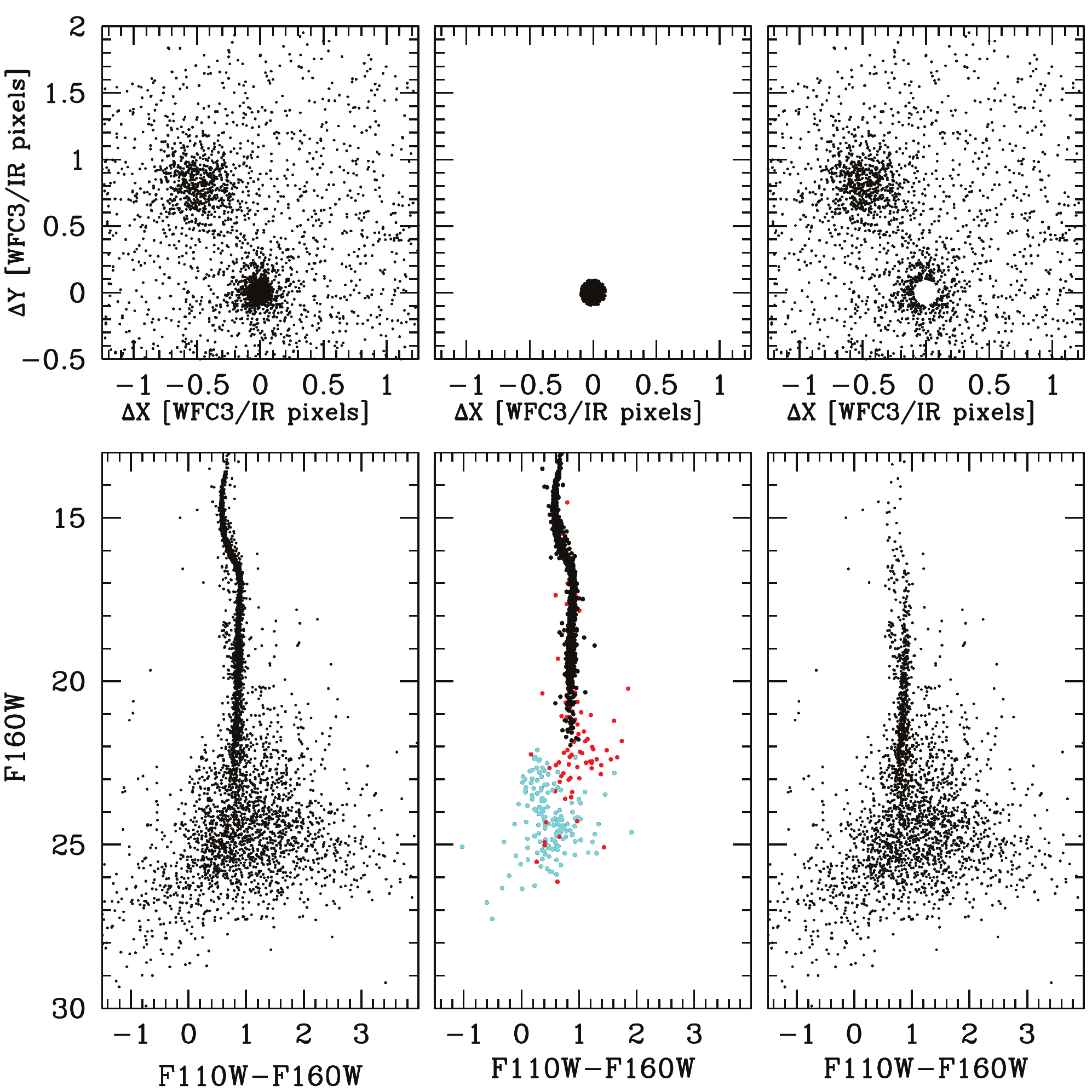}
\caption{The same as Fig.~\ref{pmcmdIR110}, but plotted for $F160W$
  vs. $F110W - F160W$.\label{pmcmdIR160}}
\end{figure*}

\begin{figure*}[h]
\centerline{
\includegraphics[height=20cm]{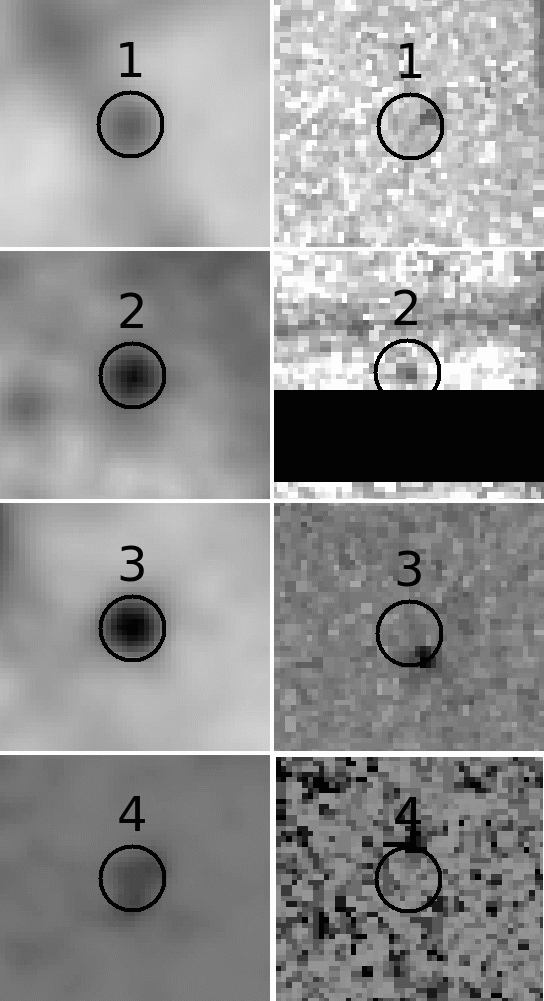}
}
\caption{Zoom on the position of the 4 faint NIR sources from
  Table~\ref{bdcand} in the stacked $F110W$ (left) and $F775W$ (right)
  images. The field of view of each image is $1.25\arcsec \times
  1.25\arcsec$ and the radius of the circles is $0.15\arcsec$. North
  is up and East to the left. \label{bdimas}}
\end{figure*}

\begin{figure*}[h]
\centerline{
\includegraphics[height=19cm]{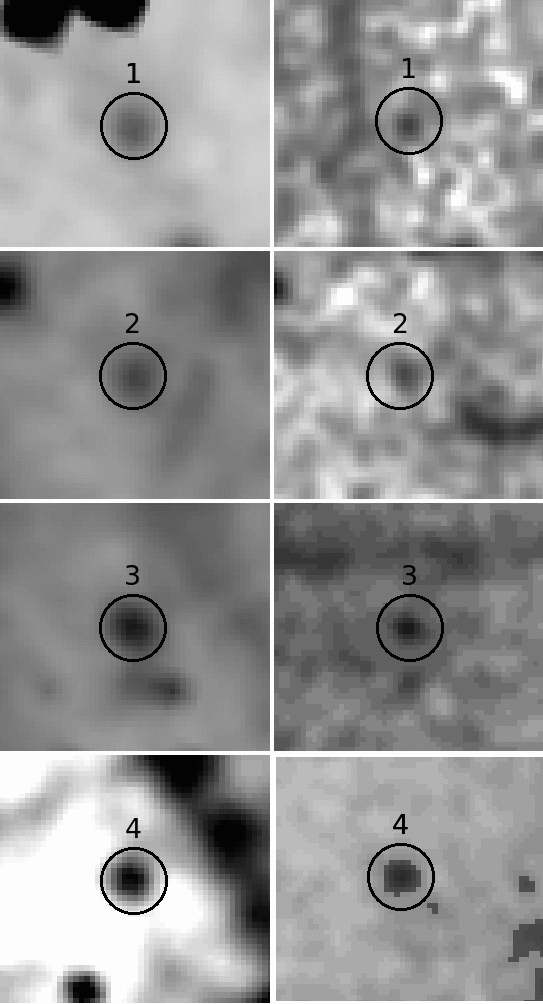}
}
\caption{Zoom on the 4 WDs from Table~\ref{wdcand} in the stacked
  $F110W$ (left) and $F775W$ (right) images. Orientation and field of
  view is the same as in Fig.~\ref{bdimas}. The WDs have similar faint
  NIR magnitudes than the BD candidates presented in
  Fig.~\ref{bdimas}, however, their optical counterpart, although
  faint, is clearly detectable. \label{wdimas}}
\end{figure*}

Figures \ref{pmcmdIR110} and \ref{pmcmdIR160} again show the NIR CMDs,
but only for those NIR sources for which an optical counterpart had
been found (black data points). The middle CMDs show only NIR sources
whose optical counterparts agree with being cluster members, based on
the displacement vector point diagram. We also show best-photometry
NIR sources {\it without} optical counterparts, plotted in red. As
mentioned in Sect.~\ref{nircmds}, all 177 best-photometry NIR sources
fainter than $F110W = 24$ mag have been inspected on the $F110W$ and
the optical $F775W$ master images. Out of these faint 177 NIR sources,
165 have an optical counterpart. Out of these 165 faint NIR/optical
sources, 48 are cluster members based on their proper motion
(displacements), and are located on the WD sequence. The remaining
five faint best-photometry NIR sources which are not on PSF streaks or
on the chip gap, and {\it without an optical counterpart} (and hence
without a proper motion estimate), all make good BD candidates to
start with. Visual inspection of these five sources and estimates on
the optical magnitude limit (see above) suggests that one is probably
a WD candidate, and the remaining four sources, listed in
Table~\ref{bdcand}, are good BD candidates.

\subsection{Expected number of BDs}

How many BDs can we expect? This number is highly uncertain and
depends on the assumed BD formation scenario (see
Sect.~\ref{intro}). Furthermore, given our detection limits we can
only expect to find the most massive BDs with masses larger than
0.068$M_\odot$ (based on the extrapolated BT-Settl models). Richer et
al.\ (2004) derived a rather flat present-day mass function for
M\,4. Extrapolating towards fainter and lower-mass stars, they
estimate that between 15 to 50 VLMSs with masses between 0.085 and
0.095 $M_\odot$ should be in their field of view (GO-8679, WFPC2
data). Using the theoretical models (Fig.~\ref{models}) we can count
the number of VLMSs in our field. If we only consider NIR sources that
have an optical counterpart and a proper motion that suggests that
they are cluster members, and best-photometry NIR sources without an
optical counterpart, we find 23 VLMSs in a mass interval between 0.08
and 0.09 $M_\odot$. Assuming that the mass function is flat and that
the slope of the mass function does not change considerably across the
stellar/sub-stellar border, we can expect a similar number of BDs with
masses between 0.070 and 0.08 $M_\odot$.  However, the number of BDs
formed per star is probably more around $\frac{1}{5}$ (see e.g. Thies
et al.\ 2007). In this case, we can expect $\approx$ 5 BDs. This is of
course a very rough estimate, but it does agree with our finding of
four BD candidates.

\subsection{Field contamination}

How many foreground or background sources can we expect in our cluster
CMD? We can simply count the number of sources that are well outside
the area covered by the cluster and the clump of field stars around
$\Delta Y \approx 0.75$ pixels and $\Delta X \approx -0.5$ pixels in
the vector point diagram. We find a field star density of 165 field
stars per $\Delta \rm{pixels}^2$. Scaling this number to the area
covered by our cluster stars, i.e., a displacement of 0.1 WFC3 pixels,
we find that we can expect 5.2 field stars in our cluster CMDs.

How many foreground stars can we then expect to have NIR magnitudes
and colours similar to our BD candidates? Selecting only field stars
with $24 < F110W < 27$ mag and $0 < F110W-F160W < 1$ mag, the field
star density is reduced to 50 stars/$\Delta \rm{pixels}^2$, and scaled
to the area covered by the cluster in the vector point diagram , we
then find that 1.6 such sources can be expected among our pm selected
faint cluster members. Thus, about half of our suggested BD candidates
might actually be foreground or background sources that happen to move
with the cluster velocity across the plane of the sky.

\section{Summary and Conclusion}
\label{summary}

We have presented the deepest NIR {\it HST}/WFC3 study of the GC M\,4
to date. The NIR data were proper-motion cleaned using archival deep
optical {\it HST}/ACS ($F775W$) data. Our best-photometry NIR CMD
reveals a narrow MS delineating down towards the expected end of the
H-burning sequence. 177 best-photometry NIR sources fainter than the
H-burning limit in $F110W$ ($F110W > 24$ mag) could be identified in
our $F110W$ master image. For 165 of these faint NIR sources, an
optical counterpart was found, 48 of these are cluster members
according to their proper motion. All of these 48 faint cluster
sources are on the WD sequence. 

We found in total five faint NIR sources for which the optical
photometry did not return a measurement (and which are not on PSF
streaks or on the chip gap). We then visually inspected the positions
of these faint NIR sources on the optical images and estimated, where
possible, upper optical magnitude limits of potential optical
counterparts that just might be visible. One source is likely another
WD and rejected as a BD candidate. Based on the upper optical
magnitude limits, we indicate the position of the remaining four
sources in the optical-NIR CMDs. One of the sources (source 4 in
Table~\ref{bdcand}), does not show an optical counterpart at all,
which implies that its optical counterpart must be fainter than the
absolute optical detection limit of $F775W > 32.9$ mag. This source
appears to be somewhat extended in the NIR image, which might indicate
multiple faint sources, i.e. multiple BDs, or possibly a
galaxy. However, its position in the CMDs does agree with this source
being a BD. One source (source 2 in Table~\ref{bdcand}) might be
another WD candidate, but its position in the optical-NIR CMD also
agrees with this source being a massive BD or a VLMS star at the
bottom of the MS. The remaining two sources also have positions that
indicate that these sources are massive BDs. We conclude that we have
found four good BD candidates, but we caution that further studies and
deeper optical data are necessary to confirm their status and cluster
membership.

\acknowledgments

We are grateful to an anonymous referee for her/his valuable comments
which helped to improve this paper. A.D.\ thanks Andrew Dolphin for
helpful discussions about {\tt DOLPHOT}. This work was supported by
NASA through grant GO-12602 from the Space Telescope Science
Institute, which is operated by AURA, Inc., under NASA contract
NAS5-26555. A.D.\ acknowledges support from the People Program (Marie
Curie Actions) of the European Union's Seventh Framework Program
FP7-PEOPLE-2013-IEF under REA grant agreement number 629579. L.R.B.\
acknowledges PRIN-INAF 2012 funding under the project entitled: "The
M4 Core Project with Hubble Space Telescope".

{\it Facilities:} \facility{HST (WFC3, ACS)}.

\end{document}